\documentclass[12pt]{article}
\usepackage{float}\restylefloat{figure}
\usepackage{epsfig}
\usepackage{graphicx}
\usepackage{epstopdf}
\def\bm#1{\mbox{\boldmath $#1$}}
\def\beq{\begin{equation}}
\def\eeq{\end{equation}}
\def\t2{\mbox{  }}
\def\rst1{\mbox{ }}

\begin{document} 
\title{\normalsize {\bf Yukawa potentials in systems with partial periodic boundary conditions I : Ewald sums for quasi-two dimensional systems.}}
\author{\normalsize Martial MAZARS\footnote{Electronic mail: Martial.Mazars@th.u-psud.fr} \\
\small Laboratoire de Physique Th\'eorique (UMR 8627),\\
\small Universit\'e de Paris Sud XI, B\^atiment 210, 91405 Orsay Cedex,
FRANCE}
\maketitle
\hfill\small\hspace{1.0in} L.P.T.-Orsay 
\begin{center}{\bf Abstract}\end{center}
Yukawa potentials are often used as effective potentials for systems as colloids, plasmas, etc. When the Debye screening length is large, the Yukawa potential tends to the non-screened Coulomb potential ; in this small screening limit, or Coulomb limit, the potential is long ranged. As it is well known in computer simulation, a simple truncation of the long ranged potential and the minimum image convention are insufficient to obtain accurate numerical data on systems. The Ewald method for bulk systems, i.e. with periodic boundary conditions in all three directions of the space, has already been derived for Yukawa potential [cf. Y., Rosenfeld, {\it Mol. Phys.\/}, \bm{88}, 1357, (1996) and G., Salin and J.-M., Caillol,  {\it J. Chem. Phys.\/}, \bm{113}, 10459, (2000)], but for systems with partial periodic boundary conditions, the Ewald sums have only recently been obtained  [M., Mazars, {\it J. Chem. Phys.\/}, {\bf 126}, 056101 (2007)]. In this paper, we provide a closed derivation of the Ewald sums for Yukawa potentials in systems with periodic boundary conditions in only two directions and for any value of the Debye length. A special attention is paid to the Coulomb limit and its relation with the electroneutrality of systems.

\newpage
\section{Introduction}

Yukawa interaction potentials, named likewise after the achievement of the meson theory by  Hideki Yukawa in 1935 \cite{f1}, were introduced by Debye and H\"uckel in 1923 as a mean field approximation in the Poisson-Boltzmann equation for ionic solution  \cite{f2} ; these potentials, in their static form, are given by
\begin{equation}
\displaystyle V(\bm{r})=Q\frac{\exp(-\kappa\mid\bm{r}\mid)}{\mid\bm{r}\mid}
\end{equation}
where $\kappa$ is the inverse screening length, and $Q$ an effective Yukawa charge. These quantities are related to some physical parameters of systems ; some examples are given in Table 1.\\
The Yukawa potential is solution of the Helmholtz equation
\begin{equation}
\displaystyle (\Delta-\kappa^2)V=-4\pi Q\delta(\bm{r})
\end{equation}
with the boundary condition $V(\infty)=0$.\\
In theories of ionic systems as liquids \cite{f3,f4}, colloids \cite{f5,f6,f6aa,f6a}, plasmas \cite{f7,f8}, etc., $\kappa^{-1}$ is known as the Debye length, this length gives an estimation of the radius of the screening sphere and $Q$ is an effective charge that may be related to the surface potential $\varphi_S$ of particles with hard-core interaction. More precisely, for hard spheres systems we have 
\begin{center}
$\displaystyle Q=\varphi_S \mbox{ } a \exp(\kappa a)$
\end{center}
with $a$ the radius of the hard spheres  (cf. Table 1).\\
In Yukawa's meson theory and according to the general principle of the quantum theory, $\kappa$ is related to the mass $m_{\pi}$ of the meson by $m_{\pi}=\kappa \hbar/c$ and $Q$ is an effective coupling constant for nuclear interactions \cite{f1}.\\
On general grounds, Yukawa interaction potentials can be considered as a reasonable approximation of effective interaction potentials between particles when some microscopic degrees of freedom may be approximated to a continuous background that screens the direct interaction between particles, while the spherical symmetry of the interaction is preserved \cite{f9}. Despite the apparent simplicity of the analytical form of the Yukawa potentials in Eq.(1), the physical mechanisms by which such effective potentials may be derived from the microscopic degree of freedom can be quite complicated and these mechanisms depend strongly on the properties of studied system. For instance, one may compare the difference between the charging mechanisms of dust grains in dusty plasma systems \cite{f7,f8,f10,f11,f11a} and the equilibrium Poisson-Boltzmann approximation or the other approximations schemes in colloidal systems \cite{f5,f6a,f9,f9a,f9b,f9c}.\\
In computer simulations, one uses periodic boundary conditions to avoid irrelevant surface bias. As the lattice sums of Yukawa potential are absolutely convergent for any value of the inverse Debye length $\kappa$, one may safely use a truncation of the potential and minimum image convention, as long as  $\kappa$ is large enough. But, in the Coulomb limit ($\kappa\rightarrow 0$), truncation of the potential introduces important bias and errors \cite{f12,f13,f14} ; therefore, there is a domain with $\kappa\neq 0$ where one has to handle lattice sums of Yukawa potential with caution by using convenient algorithm as Ewald methods \cite{f15,f16}. As outlined before by authors of refs.\cite{f15,f16}, truncation of Yukawa potentials for bulk systems may safely be used as long as $\exp(-\kappa L)/L$ is small. Typically, for systems with number density $\rho*\simeq 1$ with $N\simeq 1000$ Yukawa particles, truncation of the Yukawa potential can be used for $\kappa*=\kappa\sigma\geq 2$ (cf. Table 1) \cite{f17,f18,f19,f20}. It is also worthwhile to note that a code with Ewald sums for Yukawa potential, for any value of $\kappa$, can easily be obtained, with minimal modifications, from a code with Ewald sums for Coulomb potential \cite{f15,f16}.\\
Many interesting experimental systems in plasmas physics \cite{f21,f22} and in colloids science are quasi-two dimensional systems \cite{f22a,f23}, or even quasi-one dimensional systems \cite{f24}. To study such systems with numerical simulations, one have to use partial periodic boundary conditions in one or two directions. The purpose of the present work is to provide a detailed derivation of the Ewald method for Yukawa potentials in quasi-two dimensional systems. It seems that Ewald method for Yukawa potentials in quasi two dimensional systems have already been used in some works of refs.\cite{f22}, as outlined in ref.\cite{f22aa} ; however, the analytical details of the Ewald sums for Yukawa potentials are not gven in these works. In a paper numbered II, the Lekner sums \cite{f25,f25b} for Yukawa potentials will also be given and in the paper III, we give numerical results for a bilayer system of particles interacting via a Yukawa potential ; some preliminary results for this Yukawa bilayer system are given in the section 4 of the present paper. The analytical form of Ewald sums for Yukawa potential obtained in the present work shows that, as for bulk systems, a code can be easily written from one code for Coulomb potential with minimal modifications. In this paper, a special attention is paid to the Coulomb limit and its relation with electroneutrality. From this analysis we obtain the constants useful to compute the total energy. These constants depend on the geometry of systems and on the way the electroneutrality is achieved.\\
The present work is organised as follow. In section 2, we derive the Ewald sums of Yukawa interaction potential for quasi-two dimensional systems by using a method similar to the one used by Grzybowski and co-workers for the Coulomb interaction potential \cite{f26}. At the end of section 2, the particle-particle interaction energy for a system of $N$ particles in a quasi-two dimensional system, this result does well agree with a recent result \cite{f27} obtained from both the Ewald sums for Yukawa potential in three dimensional systems \cite{f16} and the Parry method \cite{f28}. In section 3, the small screening limit and its relations with electroneutrality of systems are outlined. In particular in this section, we show explicitly how singular contributions are cancelled by electroneutrality for three different systems ; these results allow to link the total energy of systems interacting with Yukawa potential to the energy of systems interacting with Coulomb potential. Section 4 is devoted to a discussion on the practical use of the Ewald method for Yukawa potential and to numerical tests. For completeness two appendices have been added at the end of the paper. In appendix A, we provide a full derivation of the Ewald sums for Yukawa potentials in three dimensional systems with the method by Grzybowski and co-workers \cite{f26} ; the results of this appendix agree with the derivation done by Salin and Caillol who have computed the Ewald sums from the Green's function of the Helmholtz equation \cite{f16}. In appendix B, we compute the forces obtained with the Ewald method for molecular dynamics implementations.

\section{Ewald sums for Yukawa interaction in quasi-two dimensional systems.}

We consider a system made of $N$ particles interacting via Yukawa potentials. The simulation box has partial periodic boundary conditions, with spatial periodicity $L_x$ and $L_y$, applied respectively to directions $x$ and $y$, whereas no periodic boundary conditions are taken in the third direction $z$, parallel to the unitary vector $\hat{\bm{e}}_z$. In the simulation box, the position of the particle $i$ is defined by $\bm{r}_i$ and we set 
\begin{equation}
\bm{r}_i=\bm{s}_i+z_i\mbox{ }\hat{\bm{e}}_z
\end{equation}
The particle-particle interaction energy is given by
\begin{equation}
E_{cc}(\mbox{Yukawa}; \kappa)=\frac{1}{2}\sum_{i=1}^{N}\sum_{i\neq j}^N Q_i Q_j \Phi(\bm{r}_{ij})+\frac{1}{2}\sum_{i=1}^{N}Q_i^2\Phi_0
\end{equation}
with $\Phi(\bm{r})$ defined by
\begin{equation}
\displaystyle\Phi(\bm{r})=\sum_{\bm{n}}\frac{\exp(-\kappa \mid \bm{r}+\bm{n}\mid)}{\mid \bm{r}+\bm{n}\mid}\mbox{    and    } \Phi_0\displaystyle = \sum_{\bm{n}\neq 0}\frac{\exp(-\kappa \mid \bm{n}\mid)}{\mid \bm{n}\mid}
\end{equation}
In previous equations, we have used the condensed notations 
\begin{center}
$\displaystyle \mid \bm{r}+\bm{n}\mid = \sqrt{(x+n_xL_x)^2+(y+n_yL_y)^2+z^2}$ and $\displaystyle \mid \bm{n}\mid = \sqrt{n_x^2L_x^2+n_y^2L_y^2}$
\end{center}
with $n_x$ and $n_y$ integer numbers associated with periodic images of  the box. In appendix A, one would find the computation for periodic images in all the three directions of the simulation box.\\
The computation starts with the identity 
\begin{equation}
\frac{\exp(-\kappa \mid \bm{r}+\bm{n}\mid)}{\mid \bm{r}+\bm{n}\mid} = \frac{1}{\sqrt{\pi}}\int_0^{\infty}\frac{dt}{\sqrt{t}}\exp(-\frac{\kappa^2}{4t}-\mid \bm{r}+\bm{n}\mid^2 t)
\end{equation}
that will allow to use later the Poisson-Jacobi identity.\\ 
With Eq.(6), $\Phi(\bm{r})$ may be written as
\begin{equation}
\begin{array}{ll}
\displaystyle\Phi(\bm{r})&\displaystyle=\frac{1}{\sqrt{\pi}}\int_0^{\infty}\frac{dt}{\sqrt{t}}\exp(-\frac{\kappa^2}{4t})\sum_{\bm{n}}\exp(-\mid \bm{r}+\bm{n}\mid^2 t)\\
&\\
&\displaystyle=\frac{1}{\sqrt{\pi}}\sum_{\bm{n}}\int_{\alpha^2}^{\infty}\frac{dt}{\sqrt{t}}\exp(-\frac{\kappa^2}{4t})\exp(-\mid \bm{r}+\bm{n}\mid^2 t)\\
&\\
&\displaystyle + \frac{1}{\sqrt{\pi}}\int_0^{\alpha^2}\frac{dt}{\sqrt{t}}\exp(-\frac{\kappa^2}{4t}-z^2 t)\sum_{\bm{n}}\exp(-\mid \bm{s}+\bm{n}\mid^2 t)\\
&\\
&\displaystyle = I_1(\bm{r} ; \alpha ; \kappa)+I_2(\bm{r} ; \alpha ; \kappa)
\end{array}
\end{equation}
where the integral has been split into a short ranged contribution $I_1(\bm{r} ; \alpha ; \kappa)$ and a long ranged contribution $I_2(\bm{r} ; \alpha ; \kappa)$. In Eq.(7), $\alpha$ is the Ewald damping parameter and, for practical applications, its value is chosen to balance efficiency and accuracy ; a comparison between the results of appendix A and the results in ref.\cite{f16} shows that one may interpret the parameter $\alpha$ in Eq.(7) as the gaussian width of a charge distribution that screens the central charge in the short ranged contribution [Note : In ref.\cite{f16}, the parameter $\alpha$ is denoted by $\beta$, and $\kappa$ by $\alpha$]. As remarkably outlined in ref.\cite{f16}, for three dimensional systems, the optimum choice for $\alpha$ is independent of $\kappa$ and is the same as for Ewald sums for Coulomb interaction, {\it i.e.\/} as $\kappa\rightarrow 0$.\\
With the integral relation 
\begin{equation}
\begin{array}{ll}
\displaystyle\int_{\alpha^2}^{\infty}\frac{dt}{\sqrt{t}}\exp(-\frac{\kappa^2}{4t}- a^2 t)&\displaystyle=\int_{2\alpha}^{\infty}dt\exp(-\frac{a^2t^2}{4}-{\kappa^2}{t^2})\\
&\\
&\displaystyle= \frac{\sqrt{\pi}}{2a}\mbox{\large{[}}\exp(\kappa a)\mbox{erfc}(\alpha a+\frac{\kappa}{2\alpha})+\exp(-\kappa a)\mbox{erfc}(\alpha a-\frac{\kappa}{2\alpha})\mbox{\large{]}}
\end{array}
\end{equation}
the short ranged contribution to the potential is given by
\begin{equation}
I_1(\bm{r} ; \alpha ; \kappa)=\frac{1}{2}\sum_{\bm{n}}\frac{\mbox{D}(\mid \bm{r}+\bm{n}\mid ; \alpha ; \kappa)}{\mid \bm{r}+\bm{n}\mid}
\end{equation}
where
\begin{equation}
\mbox{D}(r ; \alpha ; \kappa)=\exp(\kappa r)\mbox{erfc}(\alpha r+\frac{\kappa}{2\alpha})+\exp(-\kappa r)\mbox{erfc}(\alpha r-\frac{\kappa}{2\alpha})
\end{equation}
On Figure 1, we show the function $\mbox{D}(r ; \alpha ; \kappa)$ versus $\kappa r$ for several values of the ratio $\lambda=\kappa/\alpha$ ; on this figure, the function $\mbox{C}(r ; \alpha ; \kappa)$, used to compute the short ranged contribution of forces and defined  by Eq.(B.4) in appendix B, is also represented.\\
The long ranged contribution is computed by using the Poisson-Jacobi identity
\begin{equation}
\displaystyle\sum_{\bm{n}}\exp(-\mid \bm{s}+\bm{n}\mid^2 t) = \frac{1}{A}\frac{\pi}{t}\sum_{\bm{G}}e^{i\bm{G}.\bm{s}}\exp(-\frac{G^2}{4t})
\end{equation}
where $A=L_xL_y$ is the surface of the periodic simulation cell ; thus $I_2(\bm{r} ; \alpha ; \kappa)$ becomes
\begin{equation}
\begin{array}{ll}
\displaystyle I_2(\bm{r} ; \alpha ; \kappa)&\displaystyle=\frac{\sqrt{\pi}}{A}\int_0^{\alpha^2}\frac{dx}{x^{3/2}}\exp(-\frac{\kappa^2}{4x}-z^2x)\\
&\\
&\displaystyle + \frac{\sqrt{\pi}}{A}\sum_{\bm{G}\neq 0}  e^{i\bm{G}.\bm{s}} \int_0^{\alpha^2}\frac{dx}{x^{3/2}}\exp(-\frac{G^2+\kappa^2}{4x}-z^2x)
\end{array}
\end{equation}
With the help of Eq.(8) and after some algebra, we obtain
\begin{equation}
\begin{array}{ll}
\displaystyle \int_0^{\alpha^2}\frac{dx}{x^{3/2}}\exp(-\frac{k^2}{4x}-z^2x)&\displaystyle=\frac{\sqrt{\pi}}{k}\mbox{\large{[}}\exp(kz)\mbox{erfc}(\frac{k}{2\alpha}+\alpha z)+\exp(-kz)\mbox{erfc}(\frac{k}{2\alpha}-\alpha z)\mbox{\large{]}}\\
&\\
&\displaystyle=\sqrt{\pi} \mbox{ F}(k ; z ; \alpha)
\end{array}
\end{equation}
and thus
\begin{equation}
\displaystyle I_2(\bm{r} ; \alpha ; \kappa)=\frac{\pi}{A}\mbox{\Large{[}}\mbox{F}(\kappa;z;\alpha) + \sum_{\bm{G}\neq 0}  e^{i\bm{G}.\bm{s}} \mbox{F}(\sqrt{G^2+\kappa^2};z;\alpha) \mbox{\Large{]}}
\end{equation}
Therefore, $\Phi(\bm{r})$ is given by
\begin{equation}
\displaystyle\Phi(\bm{r})=\frac{1}{2}\sum_{\bm{n}}\frac{\mbox{D}(\mid \bm{r}+\bm{n}\mid ; \alpha ; \kappa)}{\mid \bm{r}+\bm{n}\mid}+\frac{\pi}{A}\mbox{\Large{[}}\mbox{F}(\kappa;z;\alpha) + \sum_{\bm{G}\neq 0}  e^{i\bm{G}.\bm{s}} \mbox{ F}(\sqrt{G^2+\kappa^2};z;\alpha) \mbox{\Large{]}}
\end{equation}
with $\mbox{D}(\mid \bm{r}+\bm{n}\mid ; \alpha ; \kappa)$ and $\mbox{F}(k ; z ; \alpha)$ defined respectively by Eqs.(10) and (13).\\
The lattice sums in Eq.(5) are absolutely convergent, as a consequence, as long as $\kappa\neq 0$, there is no diverging contribution in the analytical decomposition of  $\Phi(\bm{r})$ into short and long ranged contributions. As $\kappa\rightarrow 0$, a $1/\kappa$-asymptotic behaviour is found ; this stems from the conditionally convergent property of the Coulomb lattice sums. This diverging behaviour of the Coulomb lattice sums is handled by taking into account the boundary conditions as $\mid\bm{r}\mid\rightarrow\infty$ and the electroneutrality of the simulation cell \cite{f26}. The analysis of the asymptotic behaviour of Yukawa-Ewald sums as $\kappa\rightarrow 0$ and their links with the Coulomb-Ewald sums will be discussed extensively in the next section.\\
To compute the energy $E_{cc}$ of the interaction between the particles, the self contribution $\Phi_0$ is needed. This contribution is computed as for $\Phi(\bm{r})$. We have
\begin{equation}
\begin{array}{ll}
\displaystyle\Phi_0&\displaystyle = \sum_{\bm{n}\neq 0}\frac{\exp(-\kappa \mid \bm{n}\mid)}{\mid \bm{n}\mid}=\frac{1}{\sqrt{\pi}}\sum_{\bm{n}\neq 0}\int_{\alpha^2}^{\infty}\frac{dt}{\sqrt{t}}\exp(-\frac{\kappa^2}{4t})\exp(-\mid\bm{n}\mid^2 t)\\
&\\
&\displaystyle + \frac{1}{\sqrt{\pi}}\int_0^{\alpha^2}\frac{dt}{\sqrt{t}}\exp(-\frac{\kappa^2}{4t}-z^2 t)\sum_{\bm{n}}\exp(-\mid\bm{n}\mid^2 t)- \frac{1}{\sqrt{\pi}}\int_0^{\alpha^2}\frac{dt}{\sqrt{t}}\exp(-\frac{\kappa^2}{4t}-z^2 t)
\end{array}
\end{equation}
where the last integral has been added and substracted to the long ranged contribution to permit the use of the Poisson-Jacobi identity. Then
\begin{equation}
\displaystyle\Phi_0= I_1^{(0)} + I_2^{(0)}
\end{equation}
and after simple algebra we find
\begin{equation}
\left\{
\begin{array}{ll}
\displaystyle I_1^{(0)}&\displaystyle = \frac{1}{2}\sum_{\bm{n}\neq 0}\frac{\mbox{D}(\mid \bm{n}\mid ; \alpha ; \kappa)}{\mid \bm{n}\mid}\\
&\\
\displaystyle I_2^{(0)}&\displaystyle = \frac{2\pi}{A}\sum_{\bm{G}}\frac{\mbox{erfc}(\sqrt{G^2+\kappa^2}/2\alpha)}{\sqrt{G^2+\kappa^2}}-2\mbox{\Large{[}}\frac{\alpha}{\sqrt{\pi}}\exp(-\frac{\kappa^2}{4\alpha^2})-\frac{\kappa}{2}\mbox{erfc}(\kappa/2\alpha)\mbox{\Large{]}}
\end{array}
\right .
\end{equation}
With Eqs.(4), (15) and (18), the particle-particle interaction energy is given by  
\begin{equation}
\begin{array}{ll}
\displaystyle E_{cc}(\mbox{Yukawa}; \kappa)&\displaystyle=\frac{1}{4}\sum_{i, j}^NQ_iQ_j\sum_{\bm{n}}\mbox{}'\frac{\mbox{D}(\mid \bm{r}_{ij}+\bm{n}\mid ; \alpha ; \kappa)}{\mid \bm{r}_{ij}+\bm{n}\mid}+ \frac{\pi}{2A}\sum_{i, j}^NQ_iQ_j\sum_{\bm{G}\neq 0}  e^{i\bm{G}.\bm{s}_{ij}} \mbox{ F}(\sqrt{G^2+\kappa^2};z_{ij};\alpha)\\
&\\
&\displaystyle+\frac{\pi}{2A}\sum_{i, j}^NQ_iQ_j\mbox{F}(\kappa;z_{ij};\alpha)-\mbox{\Large{(}}\sum_{i=1}^{N}Q_i^2\mbox{\Large{)}}\mbox{\Large{(}}\frac{\alpha}{\sqrt{\pi}}\exp(-\frac{\kappa^2}{4\alpha^2})-\frac{\kappa}{2}\mbox{erfc}(\kappa/2\alpha)\mbox{\Large{)}}
\end{array}
\end{equation}
where the prime in the summation over the image indicates that the term $i=j$ has to be omitted for $\bm{n}=\bm{0}$.\\
The total particle-particle interaction energy is given by $E_{cc}(\mbox{Yukawa}; \kappa)$ and may be used in Monte-Carlo simulations ; Eq.(19) agrees with the results obtained by using another method in ref.\cite{f27}. For Molecular Dynamics simulations, the computation of forces is needed, this derivation is done in appendix B from the results of section 2 and 3.\\
For repulsive Yukawa interaction between particles, one needs to compute the particle-background and the background-background energies to obtain the total energy of the system and thus to be able to implement correctly a Monte-Carlo sampling of the phase space. These computations are done in the next section for three different neutralizing backgrounds, and their limits as  $\kappa\rightarrow 0$ to a non-screened Coulomb system are outlined.   

\section{The Coulomb limit and electroneutrality.}

For Coulomb interactions in systems with periodic boundary conditions, the electroneutrality property of the system is important to allow to define the electrostatic potential. Electrostatic Yukawa interactions are equivalent to non-screened Coulomb interactions as the screening length tends to infinity (or the inverse screening length $\kappa\rightarrow 0$) ; therefore, in the Coulomb limit, the Yukawa interaction energy of the system, computed with Eqs.(4-5), has to be equal to the Coulomb interaction energy. The other limit, $\kappa\rightarrow\infty$, is either a hard sphere system ($a,\sigma,\sigma_{ij} \neq 0$) or an ideal gas limit ($a,\sigma,\sigma_{ij} = 0$).\\
The series in Eq.(5) are absolutely convergent when $\kappa > 0$ (Yukawa) and conditionally convergent when $\kappa = 0$ (Coulomb). For Coulomb interactions, the conditional convergence property of the series leads to a diverging behaviour and this diverging behaviour is cancelled by using the electroneutrality of the system ; thus some constants, that depend on the geometry of the simulation box and on the manner in which electroneutrality is achieved, arise. As long as $\kappa\neq 0$, the series are absolutely convergent and no diverging behaviour is found in the total interaction energy. To achieve consistency between Yukawa and Coulomb potentials in the Coulomb limit, one has to recover from Eqs.(19) and (5) a diverging behaviour corresponding to Coulomb potential and one has to be able to cancel it by using the electroneutrality of the system. This section is devoted to the examination of the Ewald sums consistency between Yukawa and Coulomb potentials for quasi-two dimensional systems.\\
In subsection 3.1, the diverging terms appearing in the particle-particle interactions (Eq.19) as $\kappa\rightarrow 0$ are displayed and in subsections 3.2 to 3.4, we provide the results obtained for several backgrounds necessary to fulfill electroneutrality  :  a monolayer in subsection 3.2 (Fig 2.a), a bilayer (subsection 3.3, Fig 2.b) and a slab (subsection 3.4, Fig 2.c). For any other neutralizing backgrounds, one may also compute the particle-background and the background-background energies according to similar methods.

\subsection{The Particle-Particle interaction energy in the Coulomb limit.}

According to Table 1, if temperatures are high and/or if counterions are low, the inverse Debye length may become quite small. In the special case $\kappa=0$, one should recover the properties of One Component Plasma systems, thus, as $\kappa\rightarrow 0$, Coulomb interaction energy has to be recovered from Eq.(19).\\
As $\kappa\rightarrow 0$, we have
\begin{equation}
\left\{
\begin{array}{ll}
\displaystyle \mbox{D}(\mid \bm{r}\mid ; \alpha ; \kappa)&\displaystyle=2\mbox{ erfc}(\alpha \mid \bm{r}\mid)+\mbox{\Large{[}}\mbox{erfc}(\alpha \mid \bm{r}\mid)-\frac{\exp(-\alpha^2\mid \bm{r}\mid^2)}{\sqrt{\pi}\alpha\mid \bm{r}\mid}\mbox{\Large{]}}\mid \bm{r}\mid^2\kappa^2+o(\kappa^4)\\
&\\
\displaystyle\kappa\mbox{ F}(\kappa;z;\alpha)&\displaystyle=2-2\mbox{\Large{[}}z\mbox{ erf}(\alpha z)+\frac{1}{\alpha\sqrt{\pi}}e^{-\alpha^2 z^2}\mbox{\Large{]}}\kappa+\kappa^2+o(\kappa^3)
\end{array}
\right .
\end{equation}
we have also
\begin{equation}
\begin{array}{c}
G \mbox{ F}(\sqrt{G^2+\kappa^2};z;\alpha)\displaystyle=\mbox{\large{[}}\exp(Gz)\mbox{erfc}(\frac{G}{2\alpha}+\alpha z)+\exp(-Gz)\mbox{erfc}(\frac{G}{2\alpha}-\alpha z)\mbox{\large{]}}\\[0.1in]
\displaystyle + \frac{\kappa^2}{2G}\mbox{\large{[}}(1-Gz)e^{Gz} \mbox{ erfc}(\frac{G}{2\alpha}+\alpha z)-(1+Gz)e^{-Gz}\mbox{ erfc}(\frac{G}{2\alpha}-\alpha z)-\frac{2G}{\alpha\sqrt{\pi}}\exp(-\frac{G^2}{4\alpha^2}-\alpha^2z^2)\mbox{\large{]}}\\[0.1in]+o(\kappa^4)
\end{array}
\end{equation}
and 
\begin{equation}
\frac{\alpha}{\sqrt{\pi}}\exp(-\frac{\kappa^2}{4\alpha^2})-\frac{\kappa}{2}\mbox{ erfc}(\frac{\kappa}{2\alpha})=\frac{\alpha}{\sqrt{\pi}}-\frac{\kappa}{2}+\frac{3\kappa^2}{4\alpha\sqrt{\pi}}+o(\kappa^3)
\end{equation}
Therefore, we have
\begin{equation}
\displaystyle E_{cc}(\mbox{Yukawa}; \kappa\rightarrow 0)=E_{cc}(\mbox{Coulomb})+\frac{\pi}{A}\mbox{\large{(}}\sum_i Q_i\mbox{\large{)}}^2\frac{1}{\kappa}+\mbox{\large{(}}\sum_i Q_i^2\mbox{\large{)}}\frac{\kappa}{2}+o(\kappa^2)
\end{equation}
with 
\begin{equation}
\begin{array}{ll}
\displaystyle E_{cc}(\mbox{Coulomb})&\displaystyle=\frac{1}{2}\sum_{i, j}^NQ_iQ_j\sum_{\bm{n}}\mbox{}'\frac{\mbox{erfc}(\alpha\mid \bm{r}_{ij}+\bm{n}\mid)}{\mid \bm{r}_{ij}+\bm{n}\mid}+ \frac{\pi}{2A}\sum_{i, j}^NQ_iQ_j\sum_{\bm{G}\neq 0}  e^{i\bm{G}.\bm{s}_{ij}} \mbox{ F}(G;z_{ij};\alpha)\\
&\\
&\displaystyle -\frac{\pi}{A}\sum_{i, j}^NQ_iQ_j\mbox{\Large{[}}z_{ij}\mbox{ erf}(\alpha z_{ij})+\frac{1}{\alpha\sqrt{\pi}}e^{-\alpha^2 z_{ij}^2}\mbox{\Large{]}}-\frac{\alpha}{\sqrt{\pi}}\mbox{\large{(}}\sum_i Q_i^2\mbox{\large{)}}
\end{array}
\end{equation}
the Coulomb particle-particle interaction energy for a system with quasi-two dimensional geometry \cite{f26}. In Eq.(23), the singular contribution as $1/\kappa$ appears explicitly. This singular term is cancelled if the system of particles has the electroneutrality property (i.e. $\sum_i Q_i = 0$), then the small screening limit ($\kappa\rightarrow 0$) of the energy of the system, $E_{cc}(\mbox{Yukawa};\kappa)$  in Eq.(19), is exactly $E_{cc}(\mbox{Coulomb})$ in Eq.(24), the energy of the system of particles in interaction via a Coulomb potential. The electroneutrality implies that some pairs of particles in the system have attractive Yukawa interaction as, for instance, in the restricted primitive model of electrolytes (RPM) with Coulomb poyentials.\\
However, for applications to plasmas or colloids, one rather uses the Yukawa One Component Plasma model (YOCP) in which all particles have effective charge of the same sign and interactions between particles are purely repulsive ; therefore the electroneutrality can not be fulfilled by summation over the particles.  For YOCP with periodic boundary conditions in the three directions of the space, the electroneutrality is obtained by embedding the particles in a uniform neutralizing background \cite{f16} (cf. Appendix A). For quasi-two dimensional systems, such uniform volume neutralizing background can not be used since boundary conditions in the $z$-direction depend on the model studied. In the next three subsections, the small screening limits of three different neutralizing backgrounds, adapted to the quasi-two dimensional geometry, are examined. These backgrounds are monolayer, bilayer and slab neutralizing backgrounds, they are schematically represented on Figure 2.

\subsection{The monolayer neutralizing background}

The monolayer neutralizing background represented on Figure 2 (a)  is a plan with an uniform surface charge density localised at $z=0$, particles may be localised on both sides of the plan or their location may be restricted also to only one half-space. Hereafter, system (a) designs the system made by the monolayer and particles ; for this system, the charge density in the right hand side of the Helmholtz equation, Eq.(2), is given by
\begin{equation}
\displaystyle \rho_a(\bm{r})=\sum_iQ_i\delta(\bm{r}-\bm{r}_i)+\sigma\delta(z)
\end{equation}
where $\delta(z)$ is the Dirac distribution. Then, the electroneutrality reads as
\begin{equation}
NQ+A\sigma=0
\end{equation}
The energy of the system (a) is given by 
\begin{equation}
E^{(a)}(\mbox{Yukawa};\kappa)=E_{cc}(\mbox{Yukawa};\kappa)+E_{cB}(\kappa)+E_{BB}(\kappa)
\end{equation}
with
\begin{equation}
\left\{
\begin{array}{ll}
\displaystyle 
\displaystyle E_{cB}(\kappa)&\displaystyle=-\frac{NQ^2}{A}\sum_{i=1}^N\int_S d\bm{s}\int_{-\infty}^{+\infty}dz \delta(z) \sum_{\bm{n}}\frac{\exp(-\kappa \mid \bm{r}_i-\bm{r}+\bm{n}\mid)}{\mid \bm{r}_i-\bm{r}+\bm{n}\mid}\\
&\\
\displaystyle E_{BB}(\kappa)&\displaystyle=\frac{N^2Q^2}{2A^2}\int_S d\bm{s}'\int_{-\infty}^{+\infty}dz' \delta(z')\int_S d\bm{s}\int_{-\infty}^{+\infty}dz \delta(z) \sum_{\bm{n}}\frac{\exp(-\kappa \mid \bm{r}'-\bm{r}+\bm{n}\mid)}{\mid \bm{r}'-\bm{r}+\bm{n}\mid}
\end{array}
\right .
\end{equation}
$E_{cB}(\kappa)$ is the interaction energy of the particles with the background, while $E_{BB}(\kappa)$ is the self energy of the background.\\
After elementary algebra, we find 
\begin{equation}
\left\{
\begin{array}{ll}
\displaystyle 
\displaystyle E_{cB}(\kappa)&\displaystyle=-\frac{NQ^2}{A}\sqrt{\pi}\sum_{i=1}^N\int_0^{\infty}\frac{dt}{t^{3/2}}\exp(-z_i^2t-\frac{\kappa^2}{4t}) \\
&\\
\displaystyle E_{BB}(\kappa)&\displaystyle=\frac{N^2Q^2}{2A}\sqrt{\pi}\int_0^{\infty}\frac{dt}{t^{3/2}}\exp(-\frac{\kappa^2}{4t})
\end{array}
\right .
\end{equation}
and thus we have
\begin{equation}
E^{(a)}(\mbox{Yukawa};\kappa)=E_{cc}(\mbox{Yukawa};\kappa)-2\pi\frac{NQ^2}{A}\frac{1}{\kappa}\sum_{i=1}^N\exp(-\kappa\mid z_i\mid)+\pi\frac{N^2Q^2}{A}\frac{1}{\kappa}
\end{equation}
In the Coulomb limit, all singular contributions as $1/\kappa$ are cancelled by taking into account the electroneutrality (cf.Eq.(26)) ; in this limit we have
\begin{equation}
E^{(a)}(\mbox{Yukawa};\kappa\rightarrow 0)=E_{cc}(\mbox{Coulomb})+2\pi\frac{NQ^2}{A}\sum_{i=1}^N\mid z_i\mid+o(\kappa)
\end{equation}
The second contribution in Eq.(31) stems from $E_{cB}(\kappa\rightarrow 0)$. The energy of the Coulomb system is obtained by taking $\kappa=0$. The energy $E^{(a)}(\mbox{Yukawa};\kappa)$ computed with the Ewald sums is therefore defined for all value of $\kappa$ and the limit as $\kappa\rightarrow 0$ is well defined : it corresponds to a quasi-two dimensional coulombic system made of $N$ particles carrying charge $Q$ (One Component Plasma), and a monolayer with a surface charge density $\sigma$ to achieve electroneutrality, Eqs.(25) and (26). 

\subsection{The bilayer neutralizing background}

The bilayer neutralizing background is represented on Figure 2 (b), this background is made of two plans, $h$ is the separation between both plans  and $\sigma$ the surface charge density in each plan. Particles may be localised between the plans or outside the slab. We design by system (b) the system made by the bilayer and the particles. For this system, the charge density appearing in the Helmholtz equation is  
\begin{equation}
\displaystyle \rho_b(\bm{r})=\sum_iQ_i\delta(\bm{r}-\bm{r}_i)+\sigma\delta(z-\frac{h}{2})+\sigma\delta(z+\frac{h}{2})
\end{equation}
and the electroneutrality reads as
\begin{equation}
NQ+2A\sigma=0
\end{equation}
For system (b), the energy is given by 
\begin{equation}
E^{(b)}(\mbox{Yukawa};\kappa)=E_{cc}(\mbox{Yukawa};\kappa)+E_{cB}(\kappa)+E_{BB}(\kappa)
\end{equation}
with the same definitions as for system (a). With simple computations, one finds
\begin{equation}
\left\{
\begin{array}{ll}
\displaystyle 
\displaystyle E_{cB}(\kappa)&\displaystyle=-\frac{NQ^2}{A}\sqrt{\pi}\mbox{\Huge{(}}\sum_{i=1}^N\int_0^{\infty}\frac{dt}{t^{3/2}}\exp(-(z_i-\frac{h}{2})^2t-\frac{\kappa^2}{4t})\\
&\\
&\displaystyle +\sum_{i=1}^N\int_0^{\infty}\frac{dt}{t^{3/2}}\exp(-(z_i+\frac{h}{2})^2t-\frac{\kappa^2}{4t})\mbox{\Huge{)}}\\
&\\
\displaystyle E_{BB}(\kappa)&\displaystyle=\frac{N^2Q^2}{8A}\sqrt{\pi}\mbox{\Huge{(}}2\int_0^{\infty}\frac{dt}{t^{3/2}}\exp(-\frac{\kappa^2}{4t})+\int_0^{\infty}\frac{dt}{t^{3/2}}\exp(-h^2t-\frac{\kappa^2}{4t})\mbox{\Huge{)}}
\end{array}
\right .
\end{equation}
then 
\begin{equation}
\begin{array}{ll}
\displaystyle E^{(b)}(\mbox{Yukawa};\kappa)&\displaystyle=E_{cc}(\mbox{Yukawa};\kappa)-\pi\frac{NQ^2}{A}\frac{1}{\kappa}\sum_{i=1}^N\mbox{\large{(}}\exp(-\kappa\mid z_i-\frac{h}{2}\mid)+\exp(-\kappa\mid z_i+\frac{h}{2}\mid)\mbox{\large{)}}\\
&\\
&\displaystyle+\pi\frac{N^2Q^2}{2A}\frac{1}{\kappa}\mbox{\large{(}}1+\exp(-\kappa h)\mbox{\large{)}}
\end{array}
\end{equation}
Similarly to system (a), the singular contributions in system (b) are cancelled by the electroneutrality. As $h\rightarrow 0$, we recover
\begin{equation}
E^{(b)}(\mbox{Yukawa};\kappa)\rightarrow E^{(a)}(\mbox{Yukawa};\kappa)
\end{equation}
and as $\kappa\rightarrow 0$ and $h\neq 0$, one has
\begin{equation}
E^{(b)}(\mbox{Yukawa};\kappa\rightarrow 0)=E_{cc}(\mbox{Coulomb})+\pi\frac{NQ^2}{A}\sum_{i=1}^N(\mid z_i+\frac{h}{2}\mid+\mid z_i-\frac{h}{2}\mid ) -\pi\frac{N^2Q^2}{A}h+o(\kappa)
\end{equation}
The same analysis may be done easily for multilayer neutralizing backgrounds.\\
If $N_0=N/2$ particles are confined in each plan, then we have $z_i=-h/2$ or $z_i=h/2$ and the energy of the system is 
\begin{equation}
E^{(b)}(\mbox{Yukawa} ; \kappa\rightarrow 0)=E_{cc}(\mbox{Coulomb})+ 2\pi L^2\sigma^2 h +o(\kappa)
\end{equation}
then $E^{(b)}(\mbox{Yukawa};\kappa\rightarrow 0)$ is exactly the energy of the bilayer Wigner system studied in refs.\cite{f25b,f29}. This will provide an useful reference point.


\subsection{The slab neutralizing background}

The slab neutralizing background is represented on Figure 2 (c), the slab between the plans located at $z=-h/2$ and $z=h/2$ is filled with an uniform volume charge density $\rho_0$. Outside the slab, the volume charge density is null. Particles may be localised inside or outside the slab. We design by system (c) the system made by the slab and the particles. For this system, the charge density appearing in the Helmholtz equation is  
\begin{equation}
\displaystyle \rho_c(\bm{r})=\sum_iQ_i\delta(\bm{r}-\bm{r}_i)+\rho_0(\theta(z+\frac{h}{2})-\theta(z-\frac{h}{2}))
\end{equation}
with $\theta(z)$ the Heaviside distribution. For systems (c), electroneutrality reads as
\begin{equation}
NQ+Ah\rho_0=0
\end{equation}
Similarly to systems (a) and (b), the total energy of system (c) is given by 
\begin{equation}
E^{(c)}(\mbox{Yukawa};\kappa)=E_{cc}(\mbox{Yukawa};\kappa)+E_{cB}(\kappa)+E_{BB}(\kappa)
\end{equation}
with
\begin{equation}
\left\{
\begin{array}{ll}
\displaystyle 
\displaystyle E_{cB}(\kappa)&\displaystyle=-\frac{NQ^2}{Ah}\sqrt{\pi}\sum_{i=1}^N\int_{-\frac{h}{2}}^{\frac{h}{2}}dz\int_0^{\infty}\frac{dt}{t^{3/2}}\exp(-(z_i-z)^2t-\frac{\kappa^2}{4t})\\
&\\
\displaystyle E_{BB}(\kappa)&\displaystyle=\frac{N^2Q^2}{A^2h^2}\frac{4\pi}{\kappa^2}\int_S d\bm{s}'\int_{-\frac{h}{2}}^{\frac{h}{2}}dz'(1-e^{-\kappa h/2}\cosh(\kappa z'))
\end{array}
\right .
\end{equation}
To compute $E_{cB}(\kappa)$, one has to consider the set of particles that are confined in the slab and the set of particles being outside the slab, therefore we define
\begin{center}
$I=\{\mbox{  particles  } i : \mid z_i\mid > \frac{h}{2}\}$     and       $J=\{\mbox{  particles  } j : \mid z_j\mid < \frac{h}{2}\}$ 
\end{center}
and we note
\begin{center}
$\mbox{Card} I= N_I $     and      $\mbox{Card} J= N_J $ 
\end{center}
obviously we have $N_I+N_J=N$. With these definitions of sets $I$ and $J$, we find
\begin{equation}
\left\{
\begin{array}{ll}
\displaystyle E_{cB}(\kappa)&\displaystyle=-\frac{NQ^2}{Ah}\frac{\sqrt{4\pi}}{\kappa^2}\mbox{\Large{[}}\sinh(\frac{\kappa h}{2})\sum_I e^{-\kappa\mid z_i\mid}+N_J-e^{-\kappa h/2}\sum_J\cosh(\kappa z_j)\mbox{\Large{]}}\\
&\\
\displaystyle E_{BB}(\kappa)&\displaystyle=\frac{N^2Q^2}{Ah}\frac{2\pi}{\kappa^2}\mbox{\Large{[}}1-\frac{2}{\kappa h}e^{-\kappa h/2}\sinh(\frac{\kappa h}{2})\mbox{\Large{]}}
\end{array}
\right .
\end{equation}
Again, as $\kappa\rightarrow 0$, the singular contribution is cancelled by the electroneutrality and we have 
\begin{equation}
E^{(c)}(\mbox{Yukawa};\kappa\rightarrow 0)=E_{cc}(\mbox{Coulomb})+\pi\frac{N^2Q^2 h}{2L^2}\mbox{\Large{[}}\frac{N_J}{N}-\frac{1}{3}+\frac{4}{Nh}\sum_I\mid z_i \mid +\frac{4}{Nh^2}\sum_J z_j^2\mbox{\Large{]}}+o(\kappa)
\end{equation}
As $h\rightarrow 0$ and with the constraint $\rho_0 h =\sigma$, we find easily 
\begin{equation}
E^{(c)}(\mbox{Yukawa};\kappa)\rightarrow E^{(a)}(\mbox{Yukawa};\kappa)
\end{equation}

\section{Discussion}

A comparison between Eq.(19), that gives the total particle-particle interaction energy for Yukawa potential, and Eq.(24) giving the energy for Coulomb potential, shows that a code for Yukawa potential in quasi-two dimensional systems may easily be built from existing computer codes for Coulomb potential. Apart from constants, the main modifications to perform are : to change the function $\mbox{erfc}(\alpha\mid \bm{r}_{ij}+\bm{n}\mid)$ in the short range contribution to the Coulomb energy by the function $\mbox{D}(\mid \bm{r}_{ij}+\bm{n}\mid ; \alpha ; \kappa)/2$ ; in the long ranged contribution, for $\bm{G}\neq 0$, to change $G$ in the function $\mbox{ F}(G;z_{ij};\alpha)$ by $\sqrt{G^2+\kappa^2}$ and to modify the contribution for $\bm{G}=0$. When all constants arising from the backgrounds contribution and self energies are included, then we may obtain a code for Yukawa potential for any value of the inverse screening length $\kappa$, and this code will be consistent with the Coulomb limit. One should note also that the complexity of the long ranged contribution is the same as for Coulomb potential : the sum over the pairs and reciprocal lattice vectors $\bm{G}$ may not be simplified into summations over particles (with exceptions for some particular systems as the bilayers studied in refs.\cite{f29,f25b}).\\
As already noted by authors of refs.\cite{f15,f16,f18}, the use of Ewald sums for Yukawa potentials is only necessary when $\kappa$ is small, otherwise a brute truncation of the potential and the minimum image convention are sufficient to perform accurate computations. To be more specific, we may estimate a lower bound $\kappa_{cut}$ for $\kappa$ that allows to obtain a relative accuracy of the order $\delta$ in the computation of the energy. The criterion we use to estimate $\kappa_{cut}$ is as follow.\\
Consider a system of Yukawa particles with hard sphere diameter $a$, then the interaction energy at contact is $E_c=Q^2\exp(-\kappa a)/a$, while, if the separation between particles is $L/2$, with $L$ the smallest spatial periodicity of the simulation box, the energy is $E_L= 2Q^2\exp(-\kappa L/2)/L$. Then, a direct truncation of the potential will allow to obtain a relative accuracy of the order $\delta$ if $E_L/E_c < \delta$. Then, we obtain an estimation of the lower bound $\kappa_{cut}$ that allows safe truncations of Yukawa potentials for distance greater than $L/2$, as
\begin{equation}
\displaystyle \kappa^*=\kappa a > \kappa^*_{cut}=\frac{2}{L^*-2}\mbox{ ln}\mbox{\large{(}}\frac{2}{\delta L^*}\mbox{\large{)}}=\frac{\sqrt{\sigma_N^*}}{\sqrt{N}-\sqrt{\sigma_N^*}}\mbox{ ln}\mbox{\large{(}}\frac{4\sigma_N^*}{\delta^2 N}\mbox{\large{)}}
\end{equation}
where we use the reduced units $L^*=L/a$, $\kappa*=\kappa a$, $\sigma_N^*=a^2\sigma_N $ and $\sigma_N$ is the surface coverage of the simulation box defined by $\sigma_N=N/L^2$, $N$ being the number of Yukawa particles.\\
On table 2, we give some value of  $\kappa^*_{cut}$ for different values of the surface coverage and the number of particles, for relative accuracy $\delta = 10^{-4}$ and $10^{-6}$.\\
In Table 3, we report some numerical tests of the Ewald sums for Yukawa potentials. For these tests, we consider a pair of Yukawa particles with effective charge $Q_1=+1$ and $Q_2=-1$ in a box with $L_x=L_y=1$ (in these reduced units, we have $\kappa L\equiv \kappa \equiv \kappa^* L^*$) ; for this system, one has $Q_1+Q_2=0$, thus we can use straightforwardly Eq.(19) to compute the energy of the system and as $\kappa\rightarrow 0$, Eq.(23) is well behaved. The five configurations reported in Table 3 have already been considered before to test Lekner sums for Coulomb potentials \cite{f25a,f25,f29b,f29c} ; the values reported in the Column '{\it Coulomb Limit\/}' have been computed with analytical, Lekner \cite{f25} and/or the Hautman-Klein methods \cite{f29a,f29b}. For $\kappa=10^{-6}$, the results obtained with Eq.(19) perfectly agree with the Coulomb limit for Ewald damping parameters $\alpha =6$ and $\alpha =12$ and with $k\times k = 16\times 16$ for the number of vectors $\bm{G}$ in the reciprocal space. The results obtained with a direct truncation of the Yukawa potential do not allow to reproduce the correct Coulomb limit and energies computed with a direct truncation for $\kappa = 1$ are very different than the value obtained with Ewald sums. On Figure 3, we represent, for three configurations of the pair, the particle-particle energy $E_{cc}(\mbox{Yukawa}; \kappa)$ as a function of $\kappa$ (symbols) and the energy computed with a direct truncation of the Yukawa potential (lines) ; the Coulomb limit for each configuration is represented by a thick red line. It clearly appears on figure 3, that if $\kappa <10$ the use of Ewald sums is necessary ; this agrees with the results of Table 2 ($\kappa\equiv\kappa^* L^*$). One may also note on Fig.3 that, for the configuration (0.1,0.1,0.1) of the pair, by chance the energy computed with Ewald sums  agrees quite well  with the one computation using a direct truncation of the potential ; however an inspection of the numerical values shows that the difference is greater than $10^{-4}$.\\
On Figure 4, we represent  $\Delta E_{cc}(\kappa, \alpha)=  E_{cc}(\kappa, \alpha)- E_{cc}(\kappa, \alpha=6.)$ as a function of the Ewald damping parameter $\alpha$ for the three configurations of Figure 3. The contributions of the reciprocal space are computed with $k\times k = 16\times 16$ for the number of vectors $\bm{G}$. These data show the stability of the Ewald sums for Yukawa potentials versus the arbitrary choice of the damping parameter $\alpha$ ; more precisely, an inspection of the numerical data shows that, if $4.\leq \alpha < 15.$, then $\mid \Delta E_{cc}(\kappa, \alpha)\mid \leq 10^{-6}$. Therefore, a choice $\alpha L\simeq 5. - 6.$ is convenient to use Ewald sums for Yukawa potentials, as already noted for Coulomb potentials \cite{f16}. On the one hand, if $\alpha$ is too small, then the minimum image convention is insufficient and additional images have to be included in the real space contribution ; on the other hand, if $\alpha$ is too large additional $\bm{G}$-vectors have to be included in the reciprocal space contributions (cf. Fig.4).\\ 
On Table 4 and Figure 5, we report some preliminary results from Monte Carlo computations on Yukawa Bilayer systems. This system is the same as the one studied in refs.\cite{f25b,f29} but the particles interact with a repulsive Yukawa potential and the electroneutrality is achieved with the background of subsection 3.3. The geometrical and physical parameters of the bilayer correspond to run $a$ of ref.\cite{f25b} ; more precisely, we have $h=1$, $N=512$, $L=28.36$ and the coupling constant $\Gamma= Q^2/kTa=196$. Intralayer and interlayer energies and correlation functions are defined as in ref.\cite{f25b}. On Table 4, data for Coulomb interaction are extracted from ref.\cite{f25b}, the results for Yukawa interaction have been obtained with the Ewald sums given by Eqs.(19) and (26). For $\kappa=10^{-4}$, energies and correlation functions, on Fig.5 (a,b), agree perfectly ; this shows that the Coulomb limit is correctly obtained with the results of section 2 and 3. For larger values of $\kappa$, the results obtained with the Ewald sums are compared to results obtained with a direct truncation of the Yukawa potential. For $\kappa=0.1$, energies and correlation functions obtained with a direct truncation differ significantly from the results obtained with Ewald sums. For $\kappa=0.25$, energies obtained with a direct truncation of the Yukawa potential agree with energies computed with Ewald sums, but, as it appears on Fig.5 (c,d), correlation functions disagree. For larger values of $\kappa$, both energies and correlation functions  computed with Ewald sums or with a direct truncation of the interaction potential are in agreement with a good accuracy. More details and properties on Yukawa bilayer systems will be given in the forthcomming paper numbered III.\\
There are some deviations from Yukawa potentials \cite{f3,f7,f30} ; this is particularly true for quasi-two dimensional systems since counterions and coions density profiles in the $z$-directions are non-uniform. However, if, as in the theoretical analysis done in ref.\cite{f9c}, one defines an effective screening length that depends only on the $z$-coordinates of the pair of particles, then one may use the Yukawa-Ewald sums, Eq.(19), in replacing the uniform screening parameter $\kappa$ by a non-uniform $\kappa(z,z')$ (cf. Eqs.(25,26) of ref.\cite{f9c}). Otherwise, if the screening of the Coulomb interaction is more complicated, one should come back to the reliability of the Yukawa interaction potential for the system considered.\\
As a final comment to this paper, we would like to outline the fact that Coulomb limits may be obtained accurately for Yukawa potentials with the use of Ewald sums ; but, as it is shown in Section 3, the effective Coulomb limit of a system is depending on the manner the electroneutrality is restored from the background. 

\begin{center}
\large{\bf ACKNOWLEDGEMENTS}
\end{center}
I acknowledge the computation facilities of the {\it Institut du D\'eveloppement et des Ressources en Informatique Scientifique\/} (IDRIS) under the project 72104. The computations have been done on IBM Regatta Power 4. I am very grateful to D. Levesque, J.-J. Weis and J.-M. Caillol for useful and interesting discussions ; and to V. Huet  for her help in the preparation of the manuscript.

\newpage
\begin{center}
\large{\bf Appendix A: Ewald sums for Yukawa potential in three dimensional systems.}
\end{center}
\renewcommand{\theequation}{A.\arabic{equation}}
\setcounter{equation}{0}

To compute the particle-particle interaction energy of the system with periodic boundary conditions in three dimensions, we follow the same method as in section 2 for quasi-two-dimensional systems. Potentials are still defined by 
\begin{equation}
\displaystyle\Phi(\bm{r})=\sum_{\bm{n}}\frac{\exp(-\kappa \mid \bm{r}+\bm{n}\mid)}{\mid \bm{r}+\bm{n}\mid}\mbox{    and    } \Phi_0\displaystyle = \sum_{\bm{n}\neq 0}\frac{\exp(-\kappa \mid \bm{n}\mid)}{\mid \bm{n}\mid}
\end{equation}
but summations over the periodic images are taken along the three dimensions of the space. With periodic boundary conditions in all three directions of the space, the Poisson-Jacobi identity reads as
\begin{equation}
\sum_{\bm{n}}e^{-\mid \bm{r}+\bm{n}\mid^2 t}=\frac{1}{V}\mbox{\large{(}}\frac{\pi}{t}\mbox{\large{)}}^{3/2}\sum_{\bm{G}}e^{i\bm{G}.\bm{r}}\exp\mbox{\large{(}}-\frac{G^2}{4}\frac{1}{t}\mbox{\large{)}}
\end{equation}
From Eqs.(A.1) and (6), we have
\begin{equation}
\begin{array}{ll}
\displaystyle\Phi(\bm{r}) &\displaystyle=\frac{1}{\sqrt{\pi}}\sum_{\bm{n}}\int_{\alpha^2}^{\infty}\frac{dt}{\sqrt{t}}\exp(-\frac{\kappa^2}{4t})\exp(-\mid \bm{r}+\bm{n}\mid^2 t)\\
&\\
&\displaystyle + \frac{1}{\sqrt{\pi}}\int_0^{\alpha^2}\frac{dt}{\sqrt{t}}\exp(-\frac{\kappa^2}{4t}-z^2 t)\sum_{\bm{n}}\exp(-\mid \bm{r}+\bm{n}\mid^2 t)\\
&\\
&\displaystyle = I_1^{(3D)}(\bm{r} ; \alpha ; \kappa)+I_2^{(3D)}(\bm{r} ; \alpha ; \kappa)
\end{array}
\end{equation}
and with Eq.(8), we find easily 
\begin{equation}
I_1^{(3D)}(\bm{r} ; \alpha ; \kappa)=\frac{1}{2}\sum_{\bm{n}}\frac{\mbox{D}(\mid \bm{r}+\bm{n}\mid ; \alpha ; \kappa)}{\mid \bm{r}+\bm{n}\mid}
\end{equation}
with $\mbox{D}(\mid \bm{r}+\bm{n}\mid ; \alpha ; \kappa)$ given by Eq.(10). $I_2^{(3D)}(\bm{r} ; \alpha ; \kappa)$ is computed by using the Poisson-Jacobi Identity (A.2), we have then
\begin{equation}
I_2^{(3D)}(\bm{r} ; \alpha ; \kappa)=\frac{4\pi}{V}\frac{e^{-\kappa^2/4\alpha^2}}{\kappa^2}+\frac{4\pi}{V}\sum_{\bm{G}\neq 0}e^{i\bm{G}.\bm{r}}\mbox{  }\frac{\exp\mbox{\large{(}}-(G^2+\kappa^2)/4\alpha^2\mbox{\large{)}}}{(G^2+\kappa^2)}
\end{equation}
Without any difficulties, $\Phi_0$ is also computed as follows
\begin{equation}
\begin{array}{ll}
\displaystyle \Phi_0 &\displaystyle = \frac{1}{2}\sum_{\bm{n}\neq0}\frac{\mbox{D}(\mid \bm{n}\mid ; \alpha ; \kappa)}{\mid \bm{n}\mid}+ \frac{4\pi}{V}\sum_{\bm{G}\neq 0}\mbox{  }\frac{\exp\mbox{\large{(}}-(G^2+\kappa^2)/4\alpha^2\mbox{\large{)}}}{(G^2+\kappa^2)}+\frac{4\pi}{V}\frac{e^{-\kappa^2/4\alpha^2}}{\kappa^2}\\
&\\
&\displaystyle - 2\mbox{\Large{[}}\frac{\alpha}{\sqrt{\pi}}\exp(-\frac{\kappa^2}{4\alpha^2})-\frac{\kappa}{2}\mbox{ erfc}(\frac{\kappa}{2\alpha})\mbox{\Large{]}}
\end{array}
\end{equation}
Therefore the particle-particle interaction energy is given by
\begin{equation}
\begin{array}{ll}
\displaystyle E_{cc}^{(3D)}(\mbox{Yukawa}; \kappa)&\displaystyle=\frac{1}{4}\sum_{i, j}^NQ_iQ_j\sum_{\bm{n}}\mbox{}'\frac{\mbox{D}(\mid \bm{r}_{ij}+\bm{n}\mid ; \alpha ; \kappa)}{\mid \bm{r}_{ij}+\bm{n}\mid}\\
&\\
&\displaystyle +\frac{2\pi}{V}\sum_{i, j}^NQ_iQ_j\sum_{\bm{G}\neq 0}  e^{i\bm{G}.\bm{r}_{ij}}\mbox{  }\frac{e^{-(G^2+\kappa^2)/4\alpha^2}}{(G^2+\kappa^2)}+\frac{2\pi}{V}\mbox{\Large{(}}\sum_{i=1}^N Q_i\mbox{\Large{)}}^2\frac{e^{-\kappa^2/4\alpha^2}}{\kappa^2}\\
&\\
&\displaystyle -\mbox{\Large{(}}\sum_{i=1}^N Q_i^2\mbox{\Large{)}}\mbox{\Large{[}}\frac{\alpha}{\sqrt{\pi}}\exp(-\frac{\kappa^2}{4\alpha^2})-\frac{\kappa}{2}\mbox{ erfc}(\frac{\kappa}{2\alpha})\mbox{\Large{]}}
\end{array}
\end{equation}
The singular term as $\kappa\rightarrow 0$ in Eq.(A.7) is as $1/\kappa^2$, in the Coulomb limit we have
\begin{equation}
\displaystyle E_{cc}^{(3D)}(\mbox{Yukawa}; \kappa\rightarrow 0)=E_{cc}^{(3D)}(\mbox{Coulomb})+\frac{2\pi}{V}\mbox{\large{(}}\sum_i Q_i\mbox{\large{)}}^2\frac{1}{\kappa^2}+\mbox{\large{(}}\sum_i Q_i^2\mbox{\large{)}}\frac{\kappa}{2}+o(\kappa^2)
\end{equation}
with $E_{cc}^{(3D)}(\mbox{Coulomb})$ given by
\begin{equation}
\begin{array}{ll}
\displaystyle E_{cc}^{(3D)}(\mbox{Coulomb})&\displaystyle=\frac{1}{2}\sum_{i, j}^NQ_iQ_j\sum_{\bm{n}}\mbox{}'\frac{\mbox{erfc}(\alpha\mid \bm{r}_{ij}+\bm{n})}{\mid \bm{r}_{ij}+\bm{n}\mid}+ \frac{2\pi}{V}\sum_{i, j}^NQ_iQ_j\sum_{\bm{G}\neq 0}  e^{i\bm{G}.\bm{r}_{ij}} \mbox{  }\frac{e^{-G^2/4\alpha^2}}{G^2}\\
&\\
&\displaystyle -\frac{\alpha}{\sqrt{\pi}}\mbox{\large{(}}\sum_i Q_i^2\mbox{\large{)}}
\end{array}
\end{equation}
The singular term in Eq.(A.8) has to be cancelled by the electroneutrality of the system. Assuming that for all particles we have $Q_i=Q>0$ and that the electroneutrality is achieved by an uniform background $\rho_0$, then the charge density in the simulation box is 
\begin{equation}
\rho_{(3D)}(\bm{r})=\sum_iQ_i\delta(\bm{r}-\bm{r}_i)+\rho_0
\end{equation}
and the electroneutrality reads as
\begin{equation}
NQ+V\rho_0=0
\end{equation}
Thus, for the three dimensional system, the total energy is 
\begin{equation}
E^{(3D)}(\mbox{Yukawa};\kappa)=E_{cc}^{(3D)}(\mbox{Yukawa};\kappa)+E_{cB}^{(3D)}(\kappa)+E_{BB}^{(3D)}(\kappa)
\end{equation}
with
\begin{equation}
\left\{
\begin{array}{ll}
\displaystyle 
\displaystyle E_{cB}(\kappa)&\displaystyle= -\frac{NQ^2}{V}\sum_{i=1}^N\int_V d\bm{r} \sum_{\bm{n}}\frac{\exp(-\kappa \mid \bm{r}_i-\bm{r}+\bm{n}\mid)}{\mid \bm{r}_i-\bm{r}+\bm{n}\mid}\\
&\\
\displaystyle E_{BB}(\kappa)&\displaystyle=\frac{NQ^2}{2V}\int_V d\bm{r}'\int_V d\bm{r} \sum_{\bm{n}}\frac{\exp(-\kappa \mid \bm{r}'-\bm{r}+\bm{n}\mid)}{\mid \bm{r}'-\bm{r}+\bm{n}\mid}
\end{array}
\right .
\end{equation}
After a simple computation of gaussian integrals, we obtain
\begin{equation}
\left\{
\begin{array}{ll}
\displaystyle E_{cB}(\kappa)&\displaystyle= -4\pi\frac{NQ^2}{V}\frac{1}{\kappa^2}\\
&\\
\displaystyle E_{BB}(\kappa)&\displaystyle=2\pi\frac{NQ^2}{V}\frac{1}{\kappa^2}
\end{array}
\right .
\end{equation}
and therefore, for a system with an uniform neutralizing background, we have
\begin{equation}
E^{(3D)}(\mbox{Yukawa};\kappa\rightarrow 0)=E_{cc}(\mbox{Coulomb})+o(\kappa)
\end{equation}

\newpage
\begin{center}
\large{\bf Appendix B: Computation of forces.}
\end{center}
\renewcommand{\theequation}{B.\arabic{equation}}
\setcounter{equation}{0}
For molecular dynamics implementations, the computation of forces is needed ; in the following we give the forces obtained for Yukawa potentials in quasi-two dimensional systems.\\
The force acting on particle $i$ is given by
\begin{equation}
\begin{array}{ll}
\displaystyle \bm{F}_i=\bm{F}_i^{(cc)} + \bm{F}_i^{(cS)}&\displaystyle =-\bm{\nabla}_{\bm{r}_i}E_i^{(cc)}-\bm{\nabla}_{\bm{r}_i}E_i^{(cS)}\\
&\\
&\displaystyle =\sum_j \bm{F}_{j/i}^{(cc \mbox{\tiny ; short})}+\sum_j \bm{F}_{j/i}^{(cc \mbox{\tiny ;  long})}+\bm{F}_{S/i}^{(cS)}
\end{array}
\end{equation}
where $E_i^{(cc)}$ and $E_i^{(cS)}$ are respectively interaction energies of particle $i$ with other particles (cf. Eq.(19)) and with the neutralizing background (cf. Eqs.(30,36) or (43)). In Eq.(B.1), the force acting on particle $i$ due to other particles is split into a short and a long ranged contribution. With the chain rule
\begin{equation}
\displaystyle \frac{\partial f(\mid \bm{r}_{ij}+\bm{n}\mid)}{\partial x_i} =  \frac{\partial \mid \bm{r}_{ij}+\bm{n}\mid}{\partial x_i}\times\frac{d f(r)}{dr}\mbox{\Huge$\mid$}_{r=\mid \bm{r}_{ij}+\bm{n}\mid}
\end{equation}
and Eq.(19), the short range contribution of the force acting on particle $i$ due to particle $j$ is given by
\begin{equation}
\begin{array}{ll}
\displaystyle \bm{F}_{j/i}^{(cc \mbox{\tiny ; short})}&\displaystyle = -\bm{\nabla}_{\bm{r}_i}E_{j/i}^{(cc \mbox{\tiny ; short})}\\
&\\
&\displaystyle=\frac{1}{4}Q_jQ_j\sum_{\bm{n}}\mbox{}'\frac{\mbox{C}(\mid \bm{r}_{ij}+\bm{n}\mid ; \alpha ; \kappa)}{\mid \bm{r}_{ij}+\bm{n}\mid^3}\mbox{\Large{[}}(x_{ij}+n_xL_x)\hat{\bm{e}}_x+(y_{ij}+n_yL_y)\hat{\bm{e}}_y+z_{ij}\hat{\bm{e}}_z\mbox{\Large{]}}
\end{array}
\end{equation}
with $x_{ij}=x_i-x_j$ and $F(r;\alpha;\kappa)$ defined by
\begin{equation}
\displaystyle \mbox{C}(r;\alpha;\kappa)= (1-\kappa r)\exp(\kappa r)\mbox{erfc}(\alpha r+\frac{\kappa}{2\alpha})+(1+\kappa r)\exp(-\kappa r)\mbox{erfc}(\alpha r-\frac{\kappa}{2\alpha})
\end{equation}
Similarly, the long range contribution to particle-particle forces is given by
\begin{equation}
\displaystyle \bm{F}_{j/i}^{(cc \mbox{\tiny ; long})}\displaystyle = -\bm{\nabla}_{\bm{r}_i}E_{j/i}^{(cc \mbox{\tiny ; long})}=-\bm{\nabla}_{\bm{s}_i}E_{j/i}^{(cc \mbox{\tiny ; long})}-\frac{\partial}{\partial z_i}E_{j/i}^{(cc \mbox{\tiny ; long})}\hat{\bm{e}}_z
\end{equation}
From Eq.(19), we find
\begin{equation}
-\bm{\nabla}_{\bm{s}_i}E_{j/i}^{(cc \mbox{\tiny ; long})}=-\frac{\pi}{2A}Q_iQ_j\sum_{\bm{G}\neq 0} \sin(\bm{G}.\bm{s}_{ij}) \mbox{ F}(\sqrt{G^2+\kappa^2};z_{ij};\alpha)\mbox{ }\bm{G}
\end{equation}
and 
\begin{equation}
-\frac{\partial}{\partial z_i}E_{j/i}^{(cc \mbox{\tiny ; long})}=-\frac{\pi}{2A}Q_iQ_j\sum_{\bm{G}}e^{i\bm{G}.\bm{s}_{ij}} \mbox{ B}(\sqrt{G^2+\kappa^2};z_{ij};\alpha)
\end{equation}
with
\begin{equation}
\displaystyle \mbox{ B}(k;z;\alpha)=\exp(kz)\mbox{erfc}(\frac{k}{2\alpha}+\alpha z)-\exp(-kz)\mbox{erfc}(\frac{k}{2\alpha}-\alpha z)
\end{equation}
The force induced by the  background on particle $i$ depends on the geometrical parameter of the background. For systems (a), (b) and (c) described in section 3, there are some discontinuities in spatial distributions of the neutralizing background, hence forces are also discontinuous at those locations. Such behaviours may induce some complicated bias in molecular dynamics computations, especially when particles cross the surfaces of discontinuity. To implement molecular dynamics codes for systems (a), (b) and (c), one would have to estimate numerically the influence of these discontinuities on the trajectories of particles.\\
Forces induced by backgrounds on particle $i$ may be easily computed with
\begin{equation}
\displaystyle \bm{F}_{(a,b,c)/i}^{(cS)} = -\frac{\partial}{\partial z_i}E_{(a,b,c)/i}^{(cS)}\mbox{ }\hat{\bm{e}}_z
\end{equation}
For system (a), in the half-space $z_i> 0$, we have
\begin{equation}
\displaystyle \bm{F}_{(a)/i}^{(cS)} = -\frac{2\pi N Q^2}{A}\exp(-\kappa z_i)\mbox{ }\hat{\bm{e}}_z
\end{equation}
for system (b), in the slab $-h/2< z_i< h/2$,
\begin{equation}
\displaystyle \bm{F}_{(b)/i}^{(cS)} = \frac{2\pi N Q^2}{A} e^{-\kappa h/2}\sinh(\kappa z_i)\mbox{ }\hat{\bm{e}}_z
\end{equation}
and for system (c), in the slab  $-h/2< z_i< h/2$,
\begin{equation}
\displaystyle \bm{F}_{(c)/i}^{(cS)} = \sqrt{4\pi}\frac{N Q^2}{A} \frac{e^{-\kappa h/2}}{\kappa h}\sinh(\kappa z_i)\mbox{ }\hat{\bm{e}}_z
\end{equation}

\newpage
\vspace{.5cm}

\newpage
\listoftables
\normalsize{\bf Table 1 :} Definitions of the Yukawa parameters $Q$ and $\kappa$ in Eqs.(1,2) as functions of physical parameters and thermodynamical states of systems with screened-Coulomb interactions. For all systems, $k_B$ is the Boltzmann constant, $T$ the temperature, $e$ the elementary charge, $\epsilon_0$ the vacuum dielectric constant and $\epsilon$ the dielectric permittivity of the medium. In the Debye-H\"uckel approximation of electrolytes, $q_i$ and $\rho_i$ are respectively the charge and the number density of ions of species $i$, and for the Primitive Model of electrolytes,  $\sigma_{ij}=(\sigma_i+\sigma_j)/2$, is the mean ionic diameter of ions $i$ and $j$ \cite{f3,f4}. For colloidal systems, $Z$ is the charge number and $\sigma$ the hard-core diameter of macroions, $\rho$ is the number density of the monovalent counterions and $\lambda_B= e^2/\epsilon k_B T$ is the Bjerrum length \cite{f5,f6,f6aa,f6a}. In Plasmas and Dusty Plasmas, the parameters for screening-Coulomb potentials are $Z$ and $a$ the charge number and the radius of the impurity (dust particle), $n_0$ is the plasma density, $T_i$ and $T_e$ are the ion and electron temperatures \cite{f7,f8}.\\

\normalsize{\bf Table 2 :} Values of $\kappa^*_{cut}$ for several values of the surface coverage and number of particles in the simulation box. If $\kappa^*=\kappa a \geq \kappa^*_{cut}$, then one may use a direct truncation of Yukawa potential for distances greater than $L/2$ ; otherwise the long range of the potential has to be handled with a convenient algorithm.\\

\normalsize{\bf Table 3 :} Numerical tests of the Coulomb limit of the Ewald sums for the Yukawa potentials. The configuration of the pair of particles is defined by $(x_{12},y_{12},z_{12})$  particle 1 with a charge $Q_1=+1$ located at (0,0,0) and particle 2 with a charge $Q_2=-1$ located at $(x_{12},y_{12},z_{12})$ in a box with $L_x=L_y=1$. The values of particle-particle Coulomb energies are extracted from original works by others, in the original works the computations of energies have been done analytically or with the Lekner method or with the Hautmann-Klein method (HK)\cite{f29a}. Evaluations of particle-particle interaction energies in columns {\it Ewald \/}, refer to evaluations performed by using straightforwardly the Ewald sums for Yukawa potentials given by Eqs.(19), and  with $k\times k=16\times 16$ for the number of vectors $\bm{G}$ in the reciprocal space and $\alpha$ as indicated in the Table. The term {\it Truncation\/} refers to evaluations of the energies of pairs with a direct truncation at $L/2.$ of the Yukawa potentials.

\normalsize{\bf Table 4 :} Average energies computed with the Monte Carlo algorithm for a Yukawa bilayer system. The geometrical and physical parameters of the bilayer correspond to run $a$ of ref.\cite{f25b} ; the coupling constant is $\Gamma=196$, $N=512$ is the number of particles and the spatial periodicity $L=28.36$. $\beta U/N$ is the average total energy per particle, $\beta E_{intra}/N$ and $\beta E_{inter}/N$ are respectively the average intralayer and interlayer energies per particle, $\beta\sigma_{U}/N$ is an estimation of the statistical fluctuation $\sigma_U$ of the total energy. Data for the Coulomb interaction ($\kappa=0$) are extracted from ref.\cite{f25b} and data for the Direct Truncation have been computed with a direct truncation of the Yukawa potential for distance greater than $L/2$. 

\newpage 

\begin{center}
\begin{tabular}{|c|c|c|c|}
\hline
\hline
&&&\\
Systems and/or & $\kappa$ & $Q$ & physical \\
approximations & (inverse screening length) & (effective charge)  & parameters \\
&&&\\
\hline
\hline
&&&\\
Debye-H\"uckel & $\displaystyle \kappa^2=\frac{4\pi}{\epsilon k_B T} \sum_i \rho_i q_i^2$ & $\displaystyle \frac{q_i}{\epsilon}$ & $q_i$, $\rho_i$ \\
(Primitive Model &&&\\ 
of  electrolytes)&&$\displaystyle \frac{q_i\exp(\kappa\sigma_{ij})}{\epsilon (1+\kappa\sigma_{ij})}$ & and $\sigma_{ij}$\\
&&&\\
\hline
&&&\\
Colloids & $\kappa^2=4\pi\lambda_B\rho$ & $\displaystyle\frac{Ze\exp(\kappa\sigma/2)}{\epsilon(1+\kappa\sigma/2)}$ & $Z$, $\lambda_B$, $\sigma$ \\
(monovalent counterions) &&& and $\rho$ \\
&&&\\
\hline
&&&\\
Plasmas and &&&\\ 
Dusty Plasmas &&&\\ 
($a = 0$)&  $\displaystyle \kappa^2=\frac{4\pi n_0e^2}{\epsilon_0 k_B}\mbox{\large{(}}\frac{1}{T_e}+ \frac{1}{T_i}\mbox{\large{)}}$ & $\displaystyle\frac{Ze}{\epsilon_0}$ & $Z$, $n_0$, $T_i$, $T_e$ \\
&&&\\
($a\neq 0$) &&  $\displaystyle\frac{Ze\exp(\kappa a)}{\epsilon_0(1+\kappa a)}$ & and $a$ \\
&&&\\
\hline
\hline
\end{tabular}
\end{center}

\begin{center}
{\begin{quote}\item[\large\underline{\bf{Table 1}} M. Mazars, Yukawa I : Ewald sums.]\end{quote}} 
\end{center}

\newpage 

\begin{center}
\begin{tabular}{|c|ccc|ccc|}
\hline
\hline
&&&&&&\\
relative accuracy && $\delta\simeq 10^{-4}$ &&& $\delta\simeq 10^{-6}$ & \\ 
&&&&&&\\
N & 100 & 1000 & 10000 & 100 & 1000 & 10000 \\
&&&&&&\\
\hline
\hline
&&&&&&\\
$\sigma_N^*= 10$ & 15.1 & 1.9 & 0.45 & 23. & 3.1 & 0.75 \\
&&&&&&\\
$\sigma_N^*= 2$ & 3.1 & 0.7 & 0.2 & 5. & 1.2 & 0.3 \\
&&&&&&\\
$\sigma_N^*= 1$ & 2. & 0.45 & 0.1 & 3.1 & 0.75 & 0.2 \\
&&&&&&\\
$\sigma_N^*= 0.5$ & 1.2 & 0.3 & 0.07 & 2. & 0.5 & 0.15 \\
&&&&&&\\
\hline
\hline
\end{tabular}
\end{center}
\begin{center}
{\begin{quote}\item[\large\underline{\bf{Table 2}} M. Mazars, Yukawa I : Ewald sums.]\end{quote}} 
\end{center}

\footnotesize
\newpage 
\begin{center}
\begin{tabular}{|c|cc|ccc|ccc|}
\hline
\hline
&&&&&&&&\\
Configurations   &                    & Coulomb  &                  &  Ewald                        & Truncation                 &                   &Ewald              & Truncation \\ 
and References &  methods  &   Limit        & $\alpha$ &  $\kappa = 10^{-6}$ &   $\kappa = 10^{-6}$ & $\alpha$  &$\kappa =1.$ & $\kappa =1.$ \\
$(x_{12},y_{12},z_{12})$&&&&&&&&  \\
\hline
\hline
&&&&&&&&\\
(0.5,0.5,0.) & analytical & -2.28472 & 1.0  &   -1.57705  & -1.41421  & 1.0 &   -0.793469 & -0.697304     \\
\cite{f25,f29b} &           &                     & 6.0  & -2.28472    &                    & 6.0 &  -1.43067 &            \\
		       &           &                    & 12.0 & -2.28472    &  		&12.0&  -1.43067 & \\
		       &           &                    &  18.0 & -2.28481   &                    &18.0&   -1.43076 & \\
&&&&&&&&\\
(0.1,0.1,0.1) & Lekner & -5.77212     & 1.0 &  -5.74929  & -5.7735      & 1.0 &  -4.84952 & -4.85531 \\
\cite{f29b,f29c} & and   &                     & 6.0 & -5.77212  &                      & 6.0 &  -4.87041& \\
			&HK&            & 12.0& -5.77212 &                      &12.0&  -4.87041                &\\
                            &              &  & 18.0& -5.77221 &                      &18.0&   -4.8705               &\\
&&&&&&&&\\
(0.,0.,0.25)  & Lekner  & -3.72483  & 1.0   &-3.80379  & -4.0000    & 1.0 &  -3.05203 & -3.1152 \\
   \cite{f29b}  &     and        &              & 6.0  & -3.72483 &                   & 6.0 & -2.98362 & \\
		    & HK &                          &12.0 & -3.72483 &                   & 12.0 & -2.98362 &\\
		    &        &                          &18.0 & -3.72492 &                   & 18.0 & -2.98371 &\\
&&&&&&&&\\
(-0.25,-0.15,-0.2) & Lekner & -2.82156  & 1.0  & -2.73285 & -2.82843  & 1.0 &    -1.9636              & -1.98609\\
    \cite{f29b}         &     and &                   & 6.0  & -2.82156 &                   & 6.0 & -2.04483 &\\
			   & HK  &                       &12.0 & -2.82157  &                  &12.0 & -2.04483&\\
			   &         &                       &18.0 & -2.82166 &                   & 18.0 & -2.04492&\\
&&&&&&&&\\
(0.4,0.4,0.1) & Lekner & -2.28608 & 1.0 &  -1.81806 &    -1.74078    & 1.0 &  -1.03453 & -0.980076\\
  \cite{f29c}    &              &                  & 6.0 & -2.28609 &                         & 6.0 &  -1.45555 &\\
		     &              &                  &12.0 & -2.28609&                         & 12.0& -1.45555 &\\
		     &              &                  &18.0 & -2.28618 &                        & 18.0 & -1.45564 &\\
&&&&&&&&\\
\hline
\hline
\end{tabular}
\end{center}
\begin{center}
{\begin{quote}\item[\large\underline{\bf{Table 3}} M. Mazars, Yukawa I : Ewald sums.]\end{quote}} 
\end{center}

\newpage
\begin{center}
\begin{tabular}{ccccc}
\hline
\hline
&&&&\\
 $\kappa$ & $\beta U/N$ & $\beta E_{intra}/N$ & $\beta E_{inter}/N$ & $\beta\sigma_{U}/N$ \\
&&&&\\
\hline
Coulomb              & & & & \\
$\Gamma = 196$ & -219.94 & -214.87(9) & -5.06(9) & 0.046 \\
\hline
Yukawa &&&&\\
10$^{-4}$ & -219.94 & -214.87(4) & -5.07(1)    & 0.045 \\
 0.1  	    & -195.78  & -205.28(5) &  9.49(7)    & 0.045\\
 0.25          & 33.910   & -191.63(7) &  225.54(7) & 0.044 \\
2.0 		& -78.345 &  -89.52(0) &  11.17(5) & 0.058\\
&&&&\\
\hline
Direct &&&&  \\
Truncation&&&&  \\
0.1        &  602.18 & -689.60(0) & 1291.7(8) & 0.067 \\
0.25          & 33.910   & -191.63(7) &  225.54(7) & 0.044\\
2.0 & -78.343  & -89.52(2) & 11.18(0) & 0.059  \\
&&&&\\
\hline
\hline
\end{tabular}
\end{center}
\begin{center}
{\begin{quote}\item[\large\underline{\bf{Table 4}} M. Mazars, Yukawa I : Ewald sums.]\end{quote}} 
\end{center}

\newpage
{\Large\bf List of Figures}\\[0.2in]

\normalsize{\bf Figure 1:} Representation of $\mbox{D}(r ; \alpha ; \kappa)$ and  $\mbox{C}(r ; \alpha ; \kappa)$ as functions of $\kappa r$ for several values of the ratio $\lambda=\kappa/\alpha$. Function D and C are respectively defined by equations (10) and (B.4), the function D is used to compute the short ranged part of the energy while the function C is used to compute the short ranged part of the force between particles. As $\kappa r \rightarrow 0$, we have $\mbox{D}(r ; \alpha ; \kappa) \simeq 2$ and $\mbox{C}(r ; \alpha ; \kappa) \simeq 2$, this corresponds to a non screened Coulomb interaction between particles.\\ 

\normalsize{\bf Figure 2:} Schematic representation of the neutralizing backgrounds of three systems : (a) monolayer with surface charge density $\sigma=-NQ/A$, (b) bilayer with surface charge density $\sigma=-NQ/2A$ in each layer, and (c) slab with a volume charge density $\rho_0=-NQ/Ah$.\\

\normalsize{\bf Figure 3:} Representation of $E_{cc}(\mbox{Yukawa}; \kappa)$ for three configurations of a pair of Yukawa particles as functions of $\kappa$. $E_{cc}(\mbox{Yukawa}; \kappa)$ are computed by using equation (19), with $\alpha=6.0$ and $k\times k = 16\times 16$ for the number of vectors $\bm{G}$ in the reciprocal space. The particle 1 carries a charge $Q_1=+1$ and is located at (0,0,0) ; the particle 2 carries a charge $Q_2=-1$ and three positions of the particle 2 are considered in a box with periodic boundary conditions with $L_x=L_y=1.$. The three positions of particle 2 are : a : (0.1, 0.1, 0.1) ; b : (0., 0., 0.25) and c : (0.5, 0.5, 0.). Symbols refer to evaluations done with Ewald sums while lines represent evaluations done with a direct truncation at $L/2.$ of the Yukawa potentials. For each configuration, the value of the Coulomb limit is indicated by a thick horizontal line and limiting values are explicitly given. These configurations have already been considered in some previous works \cite{f25,f25a,f29b,f29c} (see also Table 3).\\

 \normalsize{\bf Figure 4:}  Dependence of the particle-particle interaction energy on the Ewald parameter $\alpha$. The configuration of the pair of particles considered are the same as those on Figure 3, with $Q_1=+1$ and $Q_2=-1$. The quantity $\Delta E_{cc}(\kappa, \alpha)=  E_{cc}(\kappa, \alpha)- E_{cc}(\kappa, \alpha=6.)$ is computed with $k\times k = 16\times 16$ for the number of vectors $\bm{G}$ in the reciprocal space and (a) $\kappa = 10^{-6}$ ; (b) $\kappa =1.$ Inspection of the numerical values shows that, if $4.\leq \alpha < 15.$, then $\mid \Delta E_{cc}(\kappa, \alpha)\mid \leq 10^{-6}$.

\normalsize{\bf Figure 5:} Representation of the intralayer $g_{11}(s)$ and interlayer $g_{12}(s)$ correlation functions obtained for the Yukawa bilayer system with the same physical parameters as in Table 4. The correlation functions are defined in ref.\cite{f25b}. For the Coulomb interaction, the correlation functions have been obtained from the ref.\cite{f25b} where the computations have been done with Ewald, Hautmann-Klein and Lekner methods.

\newpage
\begin{figure}[htbp]
\begin{center}
\centerline{\includegraphics[width=6.5in]{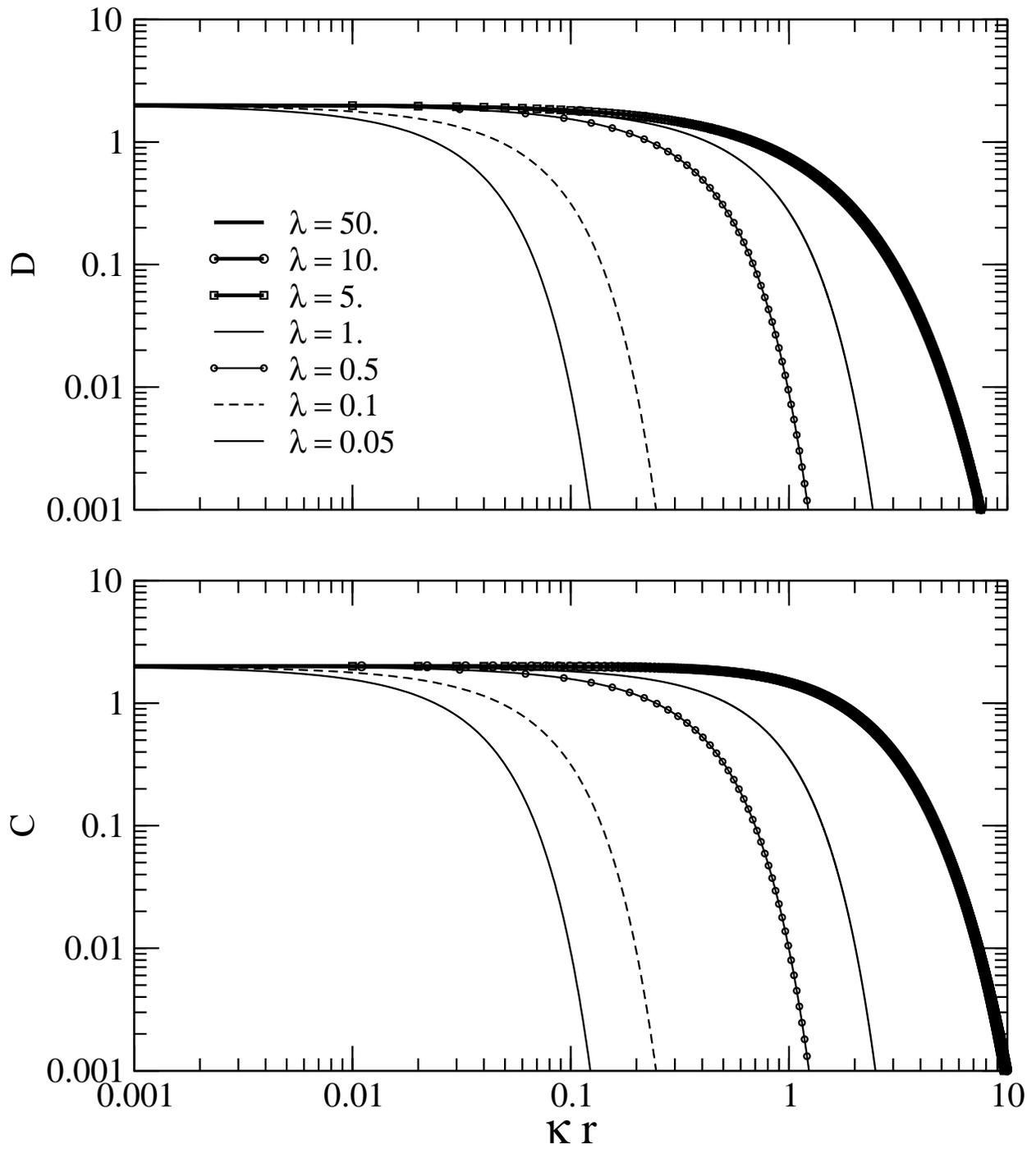}}
\caption{{\bf }  \large M. Mazars, Yukawa I : Ewald sums.}
\end{center}
\end{figure}

\newpage
\begin{figure}[htbp]
\begin{center}
\centerline{\includegraphics[width=6.5in]{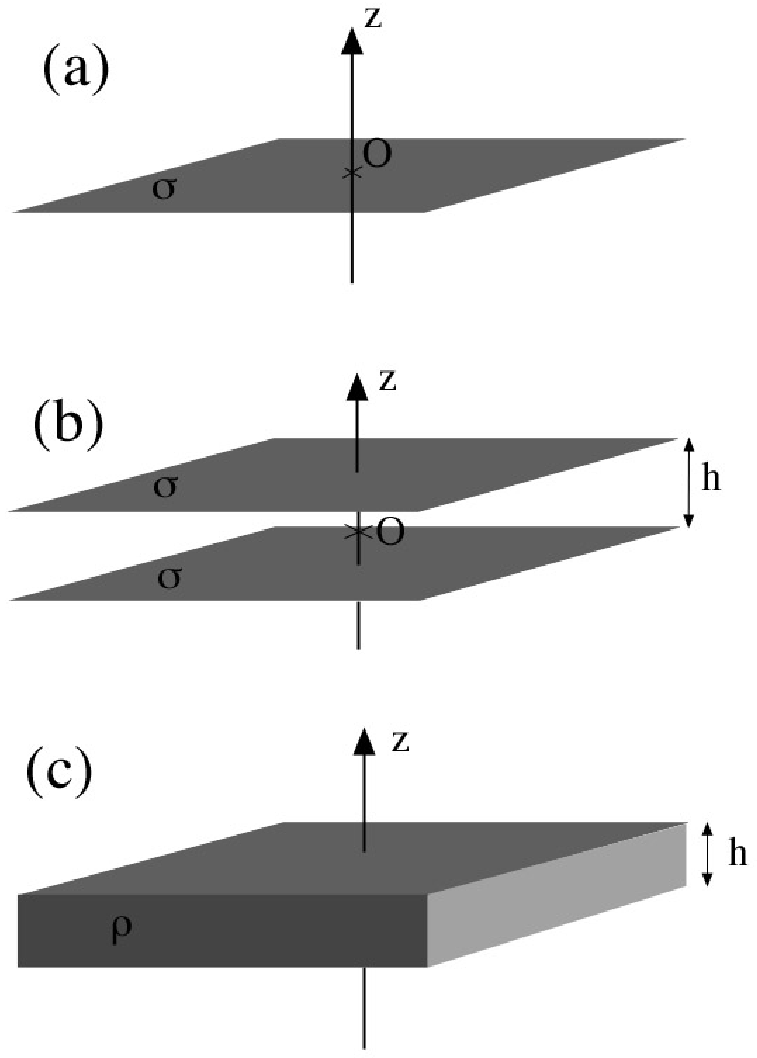}}
\caption{{\bf } \large M. Mazars, Yukawa I : Ewald sums.}
\end{center}
\end{figure}

\newpage
\begin{figure}[htbp]
\begin{center}
\centerline{\includegraphics[width=6.5in]{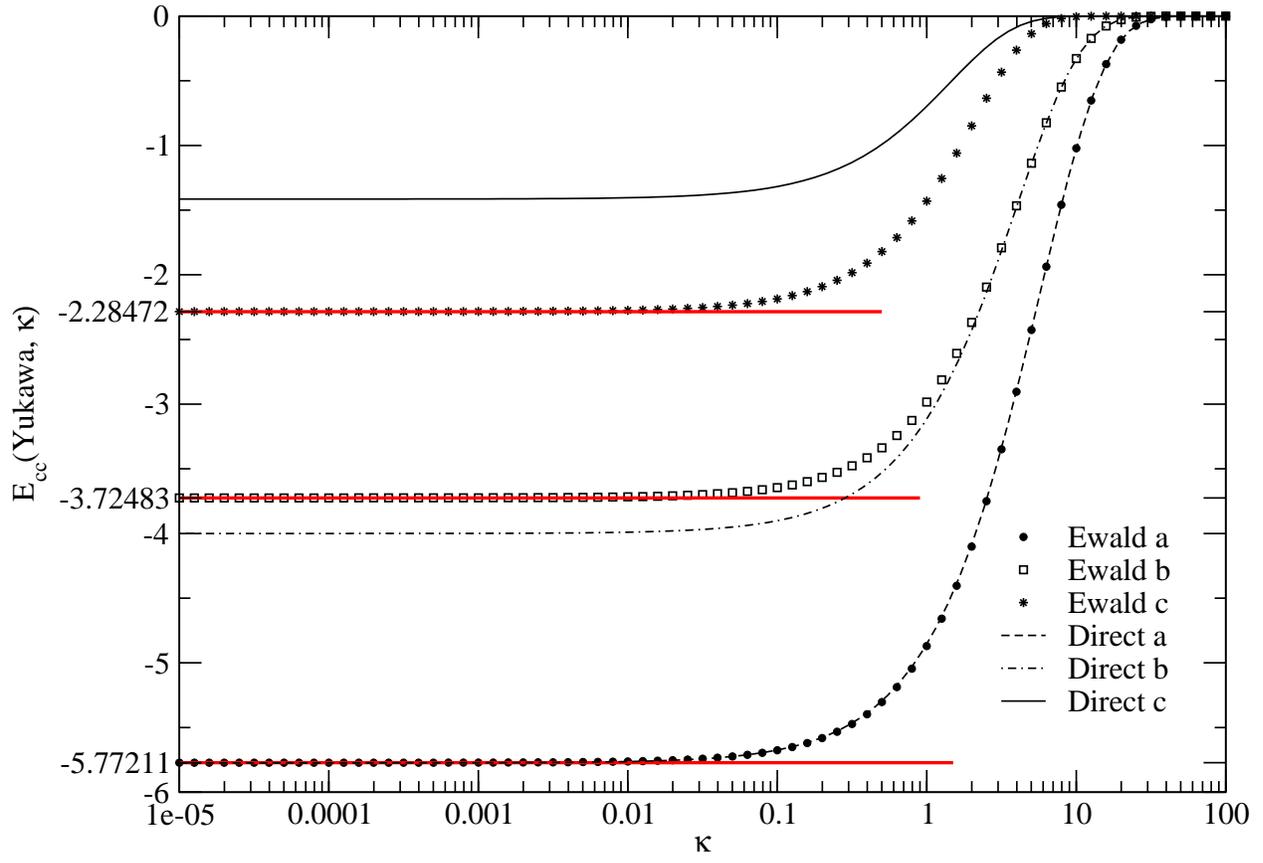}}
\caption{{\bf } \large M. Mazars, Yukawa I : Ewald sums.}
\end{center}
\end{figure}

\newpage
\begin{figure}[htbp]
\begin{center}
\centerline{\includegraphics[width=6.5in]{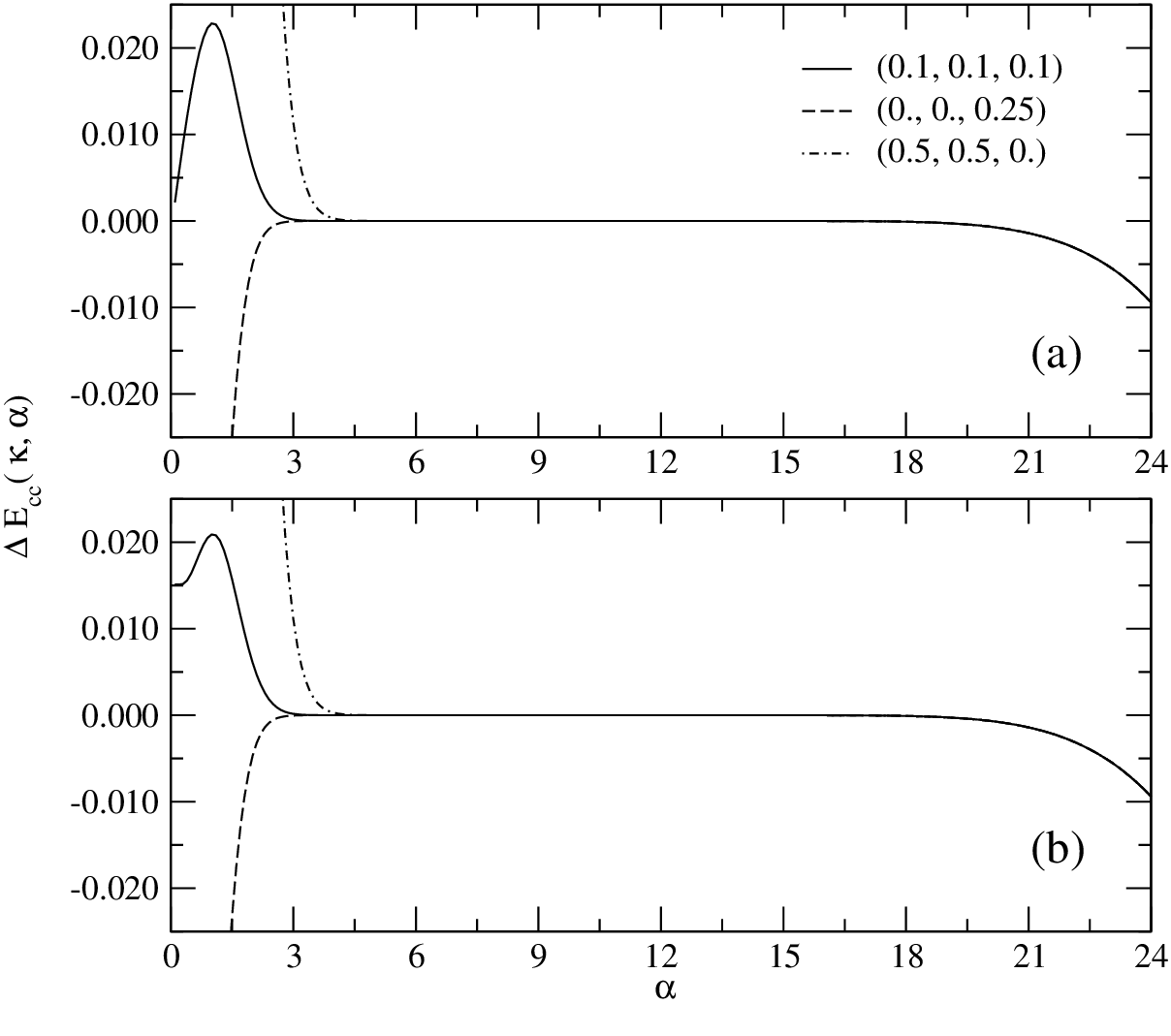}}
\caption{{\bf } \large M. Mazars, Yukawa I : Ewald sums.}
\end{center}
\end{figure}

\newpage
\begin{figure}[htbp]
\begin{center}
\centerline{\includegraphics[width=6.5in]{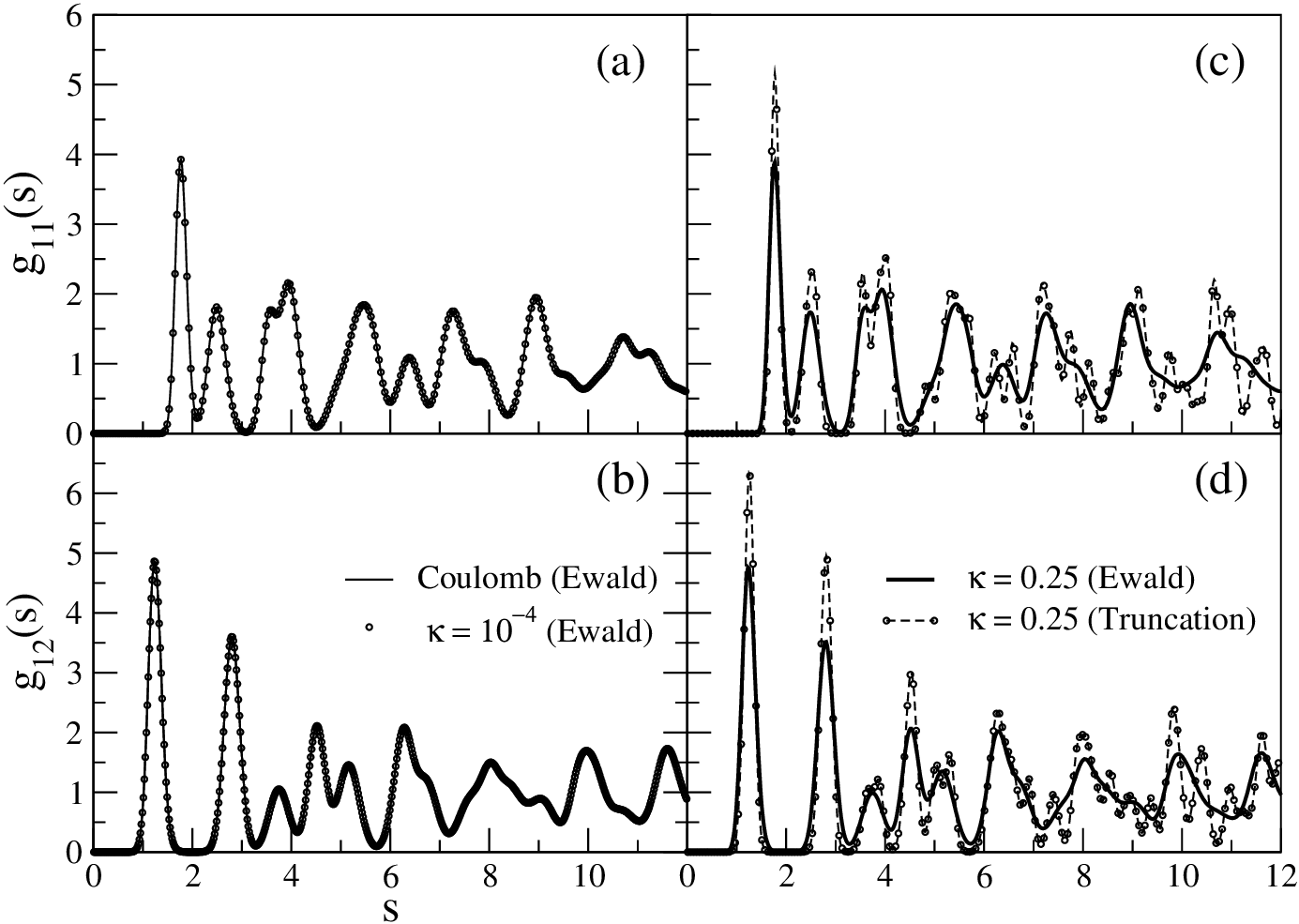}}
\caption{{\bf } \large M. Mazars, Yukawa I : Ewald sums.}
\end{center}
\end{figure}


\normalsize 
\begin{thebibliography}{99}
\bibitem{f1} H. Yukawa, {\it Meson theory in its developments.\/} Nobel lecture, December 12, 1949. (http://nobelprize.org)
\bibitem{f2} P. Debye and E. H\"uckel, {\it Phys. Z.},  {\bf 24}, 185 (1923)
\bibitem{f3} L.M. Varela, M. Garcia and V. Mosquera, {\it Phys. Rep.\/}, {\bf 382}, 1 (2003)
\bibitem{f4} M. Toda, R. Kubo and N. Sait\^o, 1992, {\it Statistical Physics I : Equilibrium Statistical Mechanics. Second Edition. (Springer).\/}
\bibitem{f5} E.J.W. Verwey and J.Th.G. Overbeek, 1999 {\it Theory of the Stability of Lyophobic Colloids. (Dover Publications, Inc.)\/} ; Originally published in 1948 by Elsevier Publishing Company, Inc. (New York).
\bibitem{f6} R.J. Hunter, 2001, {\it Foundations of Colloid Science. Second Edition.\/} (Oxford University Press)
\bibitem{f6aa} W.B. Russel, D.A. Saville and W.R. Schowalter, 1989, {\it  Colloidal Dispersions.\/} (Cambridge University Press)
\bibitem{f6a} L. Belloni, {\it J. Phys. : Condens. Matter\/}, {\bf 12}, R549 (2000)
\bibitem{f7} V.E. Fortov, A.V. Ivlev, S.A. Khrapak, A.G. Kharpak and G.E. Morfill, {\it Phys. Rep.\/}, {\bf 421}, 1 (2005)
\bibitem{f8} U. Konopka, G.E. Morfill, and L. Ratke, {\it Phys. Rev. Lett.\/}, {\bf 84}, 891 (2000)
\bibitem{f9} R. Kjellander and D.J. Mitchell, {\it Chem. Phys. Lett.},  {\bf 200}, 76 (1992)
\bibitem{f9a} R. Ramirez and R. Kjellander,  {\it J. Chem. Phys.\/}, {\bf 125}, 144110 (2006) ; Y. Han and D.G. Grier, {\it J. Chem. Phys.\/}, {\bf 122}, 064907 (2005)
\bibitem{f9b} J. Chakrabarti and  H. L\"owen, {\it Phys. Rev. E\/}, {\bf 58}, 3400 (1998)
\bibitem{f9c} A.M. Denton, and H. L\"owen, {\it Thin Solid Film\/}, {\bf 330}, 7 (1998)
\bibitem{f10} U. Konopka, L. Ratke, and H.M. Thomas, {\it Phys. Rev. Lett.\/}, {\bf 79}, 1269 (1997)
\bibitem{f11} C. Castaldo, U. de Angelis and V.N. Tsytovich, {\it Phys. Rev. Lett.\/}, {\bf 96}, 075004 (2006)
\bibitem{f11a} A.V. Filippov, A.F. Pal', A.N. Starostin, and A.S. Ivanov, {\it JETP Letters\/}, {\bf 83}, 546 (2006)
\bibitem{f12} J.C. Shelley and G.N. Patey,  Mol. Phys., {\bf 88}, 385 (1996)
\bibitem{f13} M. Patra, M. Karttunen, M.T. Hyv\"{o}nen, E. Falck, P. Lindqvist and I. Vattulainen,  Biophys. J., {\bf 84}, 3636 (2003)
\bibitem{f14} M. Bergdorf, C. Peter and P.H. H\"onenberger, J. Chem. Phys., {\bf 119}, 9129 (2003)
\bibitem{f15} Y. Rosenfeld, {\it Mol. Phys.\/}, {\bf 88}, 1357 (1996)
\bibitem{f16} G. Salin, and J.-M. Caillol, {\it J. Chem. Phys.\/}, {\bf 113}, 10459 (2000)
\bibitem{f17} G. Salin, and J.-M. Caillol, {\it Phys. Rev. Lett.\/}, {\bf 88}, 065002 (2002)
\bibitem{f18} A.-P. Hynninen, and M. Dijkstra, {\it Phys. Rev. E\/}, {\bf 68}, 021407 (2003)
\bibitem{f19} J.-M. Caillol and D. Gilles {\it J. Stat. Phys.\/}, {\bf 100}, 905 (2000) ; 933 (2000)
\bibitem{f20} T.W. Cochran and Y.C. Chiew,  {\it J. Chem. Phys.\/}, {\bf 121}, 1480 (2003)
\bibitem{f21} C.A. Knapek, D. Samsonov, S. Zhdanov, U. Konopka and G.E. Morfill, {\it Phys. Rev. Lett.\/}, {\bf 98}, 15004 (2007) ; R. Messina and H. L\"owen, {\it Phys. Rev. Lett.\/}, {\bf 91}, 146101 (2003) ; A. M. Ignatov, {\it Plasma Phys. Control. Fusion\/}, {\bf 48}, B399 (2006) ; V. Nosenko, J. Goree, and A. Piel, {\it Phys. Rev. Lett.\/}, {\bf 97}, 115001 (2006) ; T.E. Sheridan, and W.L. Therein, {\it Phys. Plasmas\/},  {\bf 13}, 062110 (2006) ; V. Nosenko, J. Goree, and A. Piel, {\it Phys. Plasmas\/},  {\bf 13}, 032106 (2006) ; V.E. Fortov, O.S. Vaulina, O.F. Petrov, {\it et al}, {\it Phys. Rev. E\/}, {\bf 75}, 026403 (2007)
\bibitem{f22}  H. Totsuji, M. Sanusi Liman, C. Totsuji, and K. Tsuruta, {\it Phys. Rev. E\/}, {\bf 70}, 016405 (2004) ;  H. Totsuji, T. Ogawa, C. Totsuji, and K. Tsuruta, {\it Phys. Rev. E\/}, {\bf 72}, 036406 (2005) ; H. Totsuji, T. Kishimoto, C. Totsuji, and T. Sasabe, {\it Phys. Rev. E\/}, {\bf 58}, 7831 (1998) ; H. Totsuji, T. Kishimoto, and K. Tsuruta, {\it Phys. Rev. Lett.\/}, {\bf 78}, 3113 (1997) ; H. Totsuji, {\it J. Phys. A: Math. Gen.\/}, {\bf 36}, 4493 (2006)
\bibitem{f22aa} H. Totsuji, T. Kishimoto, Y. Inoue, C. Totsuji and S. Nara, {\it Phys. Lett. A\/}, {\bf 221}, 215 (1996)
\bibitem{f22a} C., Bechinger, {\it Curr. Opi. Coll. \& Int. Sci.\/}, {\bf 7}, 204 (2002)
\bibitem{f23} P. Pieranski,  {\it Phys. Rev. Lett.\/}, {\bf 45}, 569 (1980) ; P. Pieranski, L. Strzelecki, and B., Pansu, {\it Phys. Rev. Lett.\/}, {\bf 50}, 900 (1983) ; C. Bechinger, Q.H. Wei and P. Leiderer, {\it J. Phys. : Condens. Matter\/}, {\bf 12}, A425 (2000) ; B. Cui, B. Lin and S.A., Rice, {\it J. Chem. Phys.\/}, {\bf 114}, 9142 (2001) ; S.H. Behrens and D.G. Grier, {\it Phys. Rev. E\/}, {\bf 64}, 050401(R) (2001) ; T. Gong and D.W.M. Marr, {\it Langmuir\/},  {\bf 17}, 2301 (2001) ; J. Santana-Solano and J.L. Arauz-Lara, {\it Phys. Rev. E\/}, {\bf 65}, 021406 (2002) ; B. Cui, H. Diamant, B. Lin and S.A. Rice, {\it Phys. Rev. Lett.\/}, {\bf 92}, 258301 (2004)
\bibitem{f24} C. Lutz, M. Kollmann, and C. Bechinger, {\it Phys. Rev. Lett.\/}, {\bf 93}, 026001 (2004) ; P.K. Yuet, {\it Langmuir\/},  {\bf 22}, 2979 (2006)
\bibitem{f25} J. Lekner, {\it Physica A\/}, {\bf 157}, 826 (1989) ; {\it Physica A\/}, {\bf 176} , 485 (1991) ; 
\bibitem{f25a} M. Mazars, {\it J. Chem. Phys.\/}, {\bf 115}, 2955 (2001)
\bibitem{f25b} M. Mazars, {\it Mol. Phys.\/}, {\bf 103}, 1241 (2005).
\bibitem{f26} E. Grzybowski, E. Gw\'o\'zd\'z,  and A. Br\'odka, {\it Phys. Rev. B\/}, {\bf 61}, 6706 (2001)
\bibitem{f27} M. Mazars, {\it J. Chem. Phys.\/}, {\bf 126}, 056101 (2007)
\bibitem{f28} D.E. Parry,  {\it Surf. Sci.\/}, {\bf 49}, 433 (1975) ; {\bf 54}, 195, (1976)
\bibitem{f29} J.-J. Weis, D. Levesque, S. Jorge. {\it  Phys. Rev. B\/}, {\bf 63}, 045308 (2001).
\bibitem{f29b} D.J. Tildesley, 1993, {\it The Molecular Dynamics Method\/}. In {\it Computer Simulation in Chemical Physics\/}, edited by M.P. Allen and D.J. Tildesley (Kluwer Academic Publishers), pp.23-47. 
\bibitem{f29c} S.Y. Liem, and J.H.R. Clarke, {\it Mol. Phys.\/}, {\bf 92}, 19 (1997)
\bibitem{f29a} J. Hautmann  and M.L. Klein, {\it Mol. Phys.\/}, {\bf 75}, 379 (1992) 
\bibitem{f30} A.R., Denton,  {\it  Phys. Rev. E\/}, {\bf 73}, 041407 (2006)

\end{thebibliography}
\end{document}